\documentclass[twocolumn,tighten,times,twocolappendix]{aastex631}

\usepackage{comment}
\usepackage{soul}

\begin{document}

\title{Variability of Central Stars of Planetary Nebulae with the Zwicky Transient Facility. I\null. Methods, Short-Timescale Variables, Binary Candidates, and the Unusual Nucleus of WeSb~1\footnote{Based in part on observations obtained with the Hobby-Eberly Telescope (HET), which is a joint project of the University of Texas at Austin, the Pennsylvania State University, Ludwig-Maximillians-Universit\"at M\"unchen, and Georg-August Universit\"at G\"ottingen. The HET is named in honor of its principal benefactors, William P. Hobby and Robert E. Eberly.}}

\shorttitle{Variability of CSPNe in ZTF}
\shortauthors{Bhattacharjee et al.}

\author[0000-0003-2071-2956]{Soumyadeep Bhattacharjee}
\affiliation{Department of Astronomy, California Institute of Technology, 1216 E. California Blvd, Pasadena, CA, 91125, USA}

\author[0000-0001-5390-8563]{S. R. Kulkarni}
\affiliation{Department of Astronomy, California Institute of Technology, 1216 E. California Blvd, Pasadena, CA, 91125, USA}

\author[0000-0002-5105-344X]{Albert K.~H. Kong}
\affiliation{Institute of Astronomy, National Tsing Hua University, Hsinchu City, Taiwan (ROC)}

\author{M. S. Tam}
\affiliation{Institute of Astronomy, National Tsing Hua University, Hsinchu City, Taiwan (ROC)}
\affiliation{Department of Physics, The Chinese University of Hong Kong, Shatin, Hong Kong}

\author[0000-0003-1377-7145]{Howard E. Bond}
\affil{Department of Astronomy \& Astrophysics, Penn State University, 
University Park, PA 16802, USA}
\affil{Space Telescope Science Institute,
3700 San Martin Dr.,
Baltimore, MD 21218, USA}

\author[0000-0002-6871-1752]{Kareem El-Badry}
\affiliation{Department of Astronomy, California Institute of Technology, 1216 E. California Blvd, Pasadena, CA, 91125, USA}

\author[0000-0002-4770-5388]{Ilaria Caiazzo}
\affiliation{Institute of Science and Technology Austria, Am Campus 1, 3400 Klosterneuburg, Austria}

\author[0000-0002-8767-3907]{Nicholas Chornay}
\affiliation{Institute of Astronomy, University of Cambridge, Madingley Road, Cambridge CB3 0HA, UK}
\affiliation{Department of Astronomy, University of Geneva, Chemin d'Ecogia 16, 1290 Versoix, Switzerland}

\author[0000-0002-3168-0139]{Matthew J. Graham}
\affiliation{Department of Astronomy, California Institute of Technology, 1216 E. California Blvd, Pasadena, CA, 91125, USA}

\author[0000-0003-4189-9668]{Antonio C. Rodriguez}
\affiliation{Department of Astronomy, California Institute of Technology, 1216 E. California Blvd, Pasadena, CA, 91125, USA}

\author[0000-0003-2307-0629]{Gregory R. Zeimann}
\affil{Hobby-Eberly Telescope, University of Texas at Austin, Austin, TX 
78712, USA}

\author[0000-0002-4223-103X]{Christoffer Fremling}
\affiliation{Department of Astronomy, California Institute of Technology, 1216 E. California Blvd, Pasadena, CA, 91125, USA}

\author{Andrew J. Drake}
\affil{Department of Astronomy, California Institute of Technology, 1216 E. California Blvd, Pasadena, CA, 91125, USA}

\author[0000-0002-6428-2276]{Klaus Werner}
\affil{Institut f\"ur Astronomie und Astrophysik, Kepler Center for
  Astro and Particle Physics, Eberhard Karls Universit\"at, Sand~1, 72076
  T\"ubingen, Germany}

\author{Hector Rodriguez}
\affil{Caltech Optical Observatories, California Institute of Technology, Pasadena, CA  91125}

\author{Thomas A. Prince}
\affil{Department of Astronomy, California Institute of Technology, 1216 E. California Blvd, Pasadena, CA, 91125, USA}

\author[0000-0003-2451-5482]{Russ R. Laher}
\affiliation{IPAC, California Institute of Technology, 1200 E. California
             Blvd, Pasadena, CA 91125, USA}

\author[0000-0001-9152-6224]{Tracy X. Chen}
\affiliation{IPAC, California Institute of Technology, 1200 E. California
             Blvd, Pasadena, CA 91125, USA}

\author[0000-0002-0387-370X]{Reed Riddle}
\affiliation{Department of Astronomy, California Institute of Technology, 1216 E. California Blvd, Pasadena, CA, 91125, USA}

\correspondingauthor{Soumyadeep Bhattacharjee}
\email{sbhatta2@caltech.edu}

%



\begin{abstract}

A complete understanding of the central stars of planetary nebulae (CSPNe) remains elusive. Over the past several decades, time-series photometry of CSPNe has yielded significant results including, but not limited to, discoveries of nearly 100 binary systems, insights into pulsations and winds in young white dwarfs, and studies of stars undergoing very late thermal pulses. We have undertaken a systematic study of optical photometric variability of cataloged CSPNe, using the light curves from the Zwicky Transient Facility (ZTF). By applying appropriate variability metrics, we arrive at a list of 94 highly variable CSPN candidates. Based on the timescales of the light-curve activity, we classify the variables broadly into short- and long-timescale variables. In this first paper in this series, we focus on the former, which is the majority class comprising 83 objects. We report periods for six sources for the first time, and recover several known periodic variables. Among the aperiodic sources, most exhibit a jitter around a median flux with a stable amplitude, and a few show outbursts. We draw attention to WeSb~1, which shows a different kind of variability: prominent deep and aperiodic dips, resembling transits from a dust\slash debris disk. We find strong evidence for a binary nature of WeSb~1 (possibly an F-type subgiant companion). The compactness of the emission lines and inferred high electron densities make WeSb~1 a candidate for either an EGB~6-type planetary nucleus, or a symbiotic system inside an evolved planetary nebula, both of which are rare objects. To demonstrate further promise with ZTF, we report three additional newly identified periodic sources that do not appear in the list of highly variable sources. Finally, we also introduce a two-dimensional metric space defined by the von Neumann statistics and Pearson Skew and demonstrate its effectiveness in identifying unique variables of astrophysical interest, like WeSb~1.

\end{abstract}




\section{Introduction} \label{sec:intro}

\defcitealias{Chornay20}{C20}
\defcitealias{Chornay21Distance}{CW21} 

The life story of most single stars in the mass range of $1-8~M_{\odot}$ is as follows: the star completes its main-sequence (MS) phase, followed by the shell-burning red-giant-branch and asymptotic-giant-branch (AGB) phases. At the end of the AGB phase, the star sheds its envelope, leaving behind the hot white dwarf at its center. The hot white dwarf ionizes the surroundings and aids in the expansion of the ejected shell through its winds (the interacting-stellar-wind mode; see \citealt{Kwok78, Frank90}). The ionized shell forms a planetary nebula (PN), and the white dwarf forms the central star of the planetary nebula (CSPN). 

This simple picture of a single star evolution, however, does not explain the non-spherical morphologies observed for most PNe \citep[e.g.,][]{Balick87, Sahai98}. This paved the way for the new understanding that many PNe result from common-envelope (CE) events in binary stellar systems \citep{Bond90Binary, Balick02, Boffin19, Jones20}. In such a binary, the primary star, as it ascends the giant branch, starts to transfer mass to the secondary. During the AGB phase, the primary has a convective envelope, so it expands as it loses mass. This process leads to the ``engulfment" of the companion by the envelope of the AGB. The envelope is ejected, primarily at the expense of the binary orbital energy. Binarity results in a spherical asymmetry, which often defines the shape of the ejected CE, perceived as a PN. Other processes like the formation of equatorial disks, and accretion jets often add further complexities \citep{Jones20}. The orbital decay of the initial binary is expected to result in a post-CE close-orbit binary inside the PN.

In the past several decades, photometric surveys have detected about a hundred binary CSPNe, thus supporting CEs as a significant formation channel for PNe. The main photometric signatures of binarity are periodic variability caused by ellipsoidal modulation of tidally deformed stars (ellipsoidal variability), irradiation of one face of a cooler companion by the hot CSPN (irradiation effect), and/or eclipses. The largest samples of binary CSPNe originated from the Optical Gravitational Lensing Experiment (OGLE; \citealt{Miszalski09a, Miszalski09b}) and {\it Kepler}-K2 \citep{Jacoby21} surveys, which comprised more than half of the known binary systems. Statistics from both these surveys indicate that $\sim$$20\%$ of CSPNe are close binaries (see the respective papers). Recently, several more binary candidates have been discovered with \emph{Gaia} \citep{Chornay21Binary, Chornay22} and the Transiting Exoplanet Survey Satellite (TESS, \citealt{Aller20, Aller24}) surveys. The orbital periods of binary CSPNe range mostly from a few hours to a few days, with a peak at around one day \citep{Jacoby21}. However, we note that periodic variability in CSPNe may alternatively result from causes other than binarity, such as rotational modulation due to starspots on a cool companion of the hot central star (see for example \citealt{Bond24}). Thus periodicity should be interpreted with caution. 

An intriguing aspect of CSPN binarity is the presence of wide binaries inside PNe. Although few in number, orbital periods ranging from $\sim$10~days to several years have been detected \citep{VanWinckel14, Jones17, Jacoby21}. This is unexpected in the picture of the binary orbital decay discussed earlier. This resulted in suggestions of alternative energy sources to eject the CE. For example, in many systems (termed grazing envelope, \citealt{Soker15,Soker20}) mass loss before CE formation (through binary interaction or winds) reduces the CE mass and subsequently the binding energy. Thus, only a small fraction of the orbital energy is needed to eject the CE, resulting in the survival of a wide binary. Recombination energy inside the envelope may also be a major (if not indispensable) energy source for unbinding the CE, reducing the burden on the orbital energy. For a concise account of these processes, see \cite{Jones20} and references therein. The survival of wide binaries inside CE is also consistent with recent discoveries of several potentially post-CE long-period white dwarf and MS binaries with \emph{Gaia} \citep{Yamaguchi24b, Yamaguchi24a}. 

Two other major sources of CSPN variability are winds from and/or pulsations of the central stars. The very young and hot white dwarfs (effective temperatures $\sim$$10^5$~K) pulsating primarily through the $\kappa$-$\gamma$ mechanism of opacity variation are named GW~Vir or PG~1159$-$035-type stars. For a review of such stars and pulsating white dwarfs in general, see \cite{Corsico19}. Several such pulsating stars have been found inside PNe (see for example \citealt{Bond90Pulsator, Ciardullo96, Hajduk14, Corsico21}). Pre-white dwarfs often possess fast winds and they can become hydrogen-deficient as a result of a late or very late thermal pulse \citep[LTP or VLTP, e.g.][]{Werner06}. Such objects exhibit spectral properties like a Wolf-Rayet star and are denoted as [WR] stars (with the brackets distinguishing them from Population~I WR stars with high-mass progenitors). For examples of such variables see \cite{Arkhipova12,Arkhipova13} and \cite{Corsico21}. Occasionally wind variability can occurs in cooler CSPNe (e.g., IC~418, \citealt{Handler97}, and similar objects).

Photometric studies of CSPNe have also led to the discovery of R Coronae Borealis (RCrB)-like variable stars inside PNe showing significant episodes of dust obscuration. Well-known systems like FG~Sge (the central star of PN Henize~1-5; \citealt{Gonzalez98}), Sakurai's Object (V4334~Sgr, central star of PN DPV~1; \citealt{Asplund97}), and V605~Aql (central star of PN Abell~58; \citealt{Clayton97}) were discovered through their unusual photometric variability. These, along with a few others (\citealt{Jeffery19}, for example), have established the (V)LTP in young (pre-)white dwarfs to be a formation channel of these objects \citep{Lawlor03}. Such discoveries have enabled the long-term monitoring of stars undergoing a (V)LTP (Sakurai's Object, for example), a very short-lived phase in the star's life, which provided valuable insights into the process.

Non-periodic photometric variability in PNe is largely an unexplored territory and there can be several other possible sources of variability. One example is irregular transits from dust/debris discs. Three such systems are PN~M~2-29 \citep{Hajduk08, Miszalski11pnm229}, NGC~2346 \citep{Mendez82}, and Hen~3-1333 \citep{Cohen02}. All of these systems show long-timescale transit-like features, conjectured to be from obscuring dust disks. A similar scenario was also recently proposed in the PN candidate PM~1-322 \citep{Paunzen23} to explain the observed unusual photometric variability in the light curve from the Zwicky Transient Facility (ZTF; \citealt{Bellm19,Graham19,Masci19,Dekany20}). The object shows prominent dips in the bluer photometric bands, associated with ``outbursts" in red bands, suggestive of dust activity. Though the object is now thought to be a symbiotic system, confirmatory evidence of its PN or symbiotic nature has not been obtained and it remains a system of interest. 

The above discussion establishes the importance of a systematic study of the photometric variability of CSPNe. Recently, \cite{Chornay21Binary} used the \emph{Gaia}-EDR3 photometric data to conduct such a study. They arrived at a list of highly variable CSPNe, several of which are likely candidates for binary systems. In this series of two papers, we perform a similar study but using the epochal photometric data from ZTF. By using suitable variability metrics, we shortlist the most significantly variable CSPNe and focus on them, uncovering several different classes of variables. Based on the timescale of variability, we broadly classify our objects into short- and long-timescale variables. In Paper I (this work), we focus on the former and discuss the latter in Paper II (Bhattacharjee et al. in prep.). Among the variable sources, a few stand out with their highly unusual light curves, which we discuss in further detail. In this paper, we choose WeSb~1 as a case study.

The paper is broadly organized into two parts. The first part, comprising the next two sections, deals with the methodology and the main results. In Section \ref{sec:method} we describe the methodology of CSPN selection, photometric quality cuts, and the variability metrics. In Section \ref{sec:results}, we discuss the results and compare them with previous works. Section \ref{appendix:wesb1} comprises the second part of the paper, where we present our analysis of WeSb~1.  Finally we present our conclusions in Section \ref{sec:conclusion}.

\section{Methods}\label{sec:method}

\subsection{Starting Sample}

The largest sample of confirmed and candidate PNe is publicly available in the Hong Kong/Australian Astronomical Observatory/Strasbourg Observatory ${\rm H\alpha}$ Planetary Nebula (HASH) catalog \citep{Parker16}. Recently, \citet[][hereafter \citetalias{Chornay21Distance}]{Chornay21Distance} used \emph{Gaia}-EDR3 colors and parallaxes to identify the central stars of the HASH PNe and assigned reliability to the identifications (for the algorithm, see their earlier paper based on \emph{Gaia}-DR2: \citealt{Chornay20}). 

We base our study on this \emph{Gaia}-EDR3-selected CSPNe of \citetalias{Chornay21Distance}. We limit our sample to those with $>$50\% CSPN identification reliability (as was also done in \citealt{Chornay21Binary}). With this cut, our starting sample has 1812 high-confidence CSPNe. Since the work of \citetalias{Chornay21Distance}, the classifications of a few objects have been modified in the HASH catalog. We do not account for this in our selection. However, in our final tabulation of variable objects, we include the updated PN status from the current version of the HASH catalog. 

\subsection{Photometry and Associated Problems}\label{subsec:photometry}

ZTF is a 47\,deg$^2$ field-of-view camera attached to the 48-inch Samuel Oschin telescope at Palomar Observatory. In operation since 2018 March, ZTF phase I (2018 March -- 2020 Sep) observed the entire northern sky in $g$- and $r$-bands at declinations $\delta>-30\,$deg with a three-night cadence, while ZTF phase II (2020 Oct -- 2023 Sep) and ZTF - O4 (2023 Sept -- 2024 Sept) observed with a two-night cadence. It operates in primarily $g$ (4000 -- 5500~\AA) and $r$ (5500 -- 7200~\AA) photometric bands. Observations in the $i$-band (7000 -- 8500~\AA) are performed outside the main survey in partnership programs. Thus, the coverage in this band is often insufficient. For pointing, the sky is divided into about 700 slightly overlapping primary fields, which cover nearly $\sim$$90\%$ of the survey footprint. A similar number of secondary fields are also used to at least partially compensate for the area loss resulting from CCD gaps in the primary fields. Thus, an object can be observed under multiple ZTF fields.

The data can be accessed from two sources employing two different photometric methodologies. The first is the photometry on the science exposures, which relies on the point-spread function (PSF) source identification in each image (we call this the standard photometry). These data are available through the NASA/IPAC Infrared Science Archive (IRSA) database.\footnote{\url{https://irsa.ipac.caltech.edu/Missions/ztf.html}} The second uses forced PSF photometry on difference images at user-input coordinates (we refer to this as difference photometry). For each field, around 15 -- 20 exposures are used to build a static reference image, which is subtracted from each exposure to construct the difference images (see \citealt{Masci23}). With extended sources, such as PNe, the standard photometry can be erroneous as PSF-source identification is not well-defined. Thus, we primarily use the difference photometry at the \emph{Gaia} coordinates of the CSPNe. The underlying assumption is that the nebular flux is static over the extent of the CSPN, which gets subtracted out in the difference images.

We generated the ZTF light curves using the batch forced-photometry service, described in \citet{Masci23}. Photometric data could be successfully generated for 991 of the objects. This is slightly less than that expected solely from the ZTF sky coverage fraction. The extra data loss results from several internal quality checks in the difference photometry pipeline. In the variability study, the quantity of interest is the fractional variation of flux. To obtain this, we convert the difference flux to absolute flux and magnitude, following the prescription laid out in \citet{Masci23}.

In the process, we found two main issues. The first is step discontinuities in the difference fluxes in a few ($\sim$$10$) objects. Manual inspection shows that this artifact likely affected only the last two years of ZTF data and is possibly limited to one CCD (bearing ID number 4, though this could not be conclusively established). Not all the objects in the field are uniformly affected by this problem, which prevents normalization as a solution. We checked that the standard photometry is not affected by this, which indicates that the issue is specific to difference photometry. The reason behind this artifact is yet unclear and may result from either abrupt changes in the reference image and/or difference image generation pipeline. 

The second issue pertains to the inference of the reference fluxes from the reference images (which are used to obtain the absolute fluxes from the difference fluxes). Possibly due to the uncertainties with photometry on extended sources, for a few objects, the reference fluxes from the different ZTF fields do not agree with each other (disagreement of $\sim$$0.2-0.5$ magnitudes, thus significant). Visually, these mostly are objects with very bright and extended nebulae, thus prominently extended sources (without a resolved CSPN). As with the previous issue, a simple solution was not found.

To avoid the significant contamination of these artifacts in our variable sample selection, we adopt the following workaround solutions. In addition to the difference photometry, we also use the light curves from standard photometry. We searched a $1''$-radius cone centered on the \emph{Gaia} coordinates using {\tt ztfquery} \citep{mickael_rigault_2018_1345222}. We apply the variability metrics to both the forced and standard photometry data and use a joint cut on the metric value to define our variable sample. The process is described in section \ref{subsec:metrics_and_period}. This mitigates the first issue. To eliminate objects significantly affected by the second issue, we employ an additional post-processing step with the difference photometry data. This step is described in Section \ref{subsec:quality_cuts}. 

We note here that the above procedures aimed at eliminating the photometric artifacts result in significant data loss and heightened error bars. This leaves ample scope for improvement. For example, a simple solution to the first issue could be to use the \emph{Gaia} magnitudes to obtain the reference fluxes. Such an exercise, however, introduces additional uncertainties on several fronts (for example, the difference in photometric band passes and observation times) which need to be quantified. We also note that the current ZTF photometry uses the nightly PSFs generated by the algorithm described in \cite{Masci19}. Changing PSF may thus result in varying amounts of nebular flux inclusion which may result in artificial variability. We expect this to be an issue primarily with the standard photometry (as, in difference photometry, the nebula gets adequately subtracted in most cases), thus using both photometries mitigates this effect. But we do not rule out a few contaminants arising from this effect in our subsequent samples. In the long run, a reliable photometric method for PNe (and, in general, for any extended source with embedded stars) needs to be formulated.

\subsection{Quality Cuts and Post-processing}\label{subsec:quality_cuts}

\textit{Difference Image Photometry:} We closely follow the quality cuts prescribed in \citet{Masci23}. These are devised mainly for point sources and may not be strictly applicable to extended objects like PNe. However, although at the cost of completeness, these ensure that the sources have reliable ZTF light curves. We require that the distance to the nearest reference source (used to generate the reference flux), \texttt{dnearestrefsrc}, is $<$$1''$ from the \emph{Gaia} CSPN coordinate. We also reject all data points with \texttt{procstatus}$\neq$0, \texttt{infobitssci}$\neq$0, and \texttt{scisigpix}$>$25, which govern the proper execution of the difference photometry and associated pixel noise. Data points obtained at seeing, \texttt{sciinpseeing}$\geq$4$''$ were also rejected. To ensure that the reference source is not too far from a point source (thus ensuring that the PSF photometry is sufficiently trustworthy), we restrict the sharp and chi parameters based on the PSF-fit to be in the ranges of \texttt{nearestrefchi}$<$$2.5$ and $-0.25$$<$\texttt{nearestrefsharp}$<$$0.25$. Additionally, we reject all data points for which any of the quantities has a null value. With these quality cuts, most of the sources with `problematic' ZTF light curves are rejected. Nevertheless, we perform the additional post-processing step to tackle the second issue described in section \ref{subsec:photometry}. Note that the results are not significantly affected with or without this step.

Firstly, we reject all secondary field images (i.e., fields with IDs, \texttt{field}$\geq$$1000$). This is because the reference images for the secondary fields are often either not present or not well-recorded (recorded over half a CCD, or at a different time of year than the primary fields). Following this, we take the mean of the reference magnitude readings (\texttt{nearestrefmag}) across all the ZTF fields and use it as a final reference magnitude. This combination is reasonable with the primary ZTF fields as the reference images were recorded over similar timelines. To incorporate the additional uncertainty introduced in taking the mean, we add the standard deviation of the \texttt{nearestrefmag} readings to the corresponding uncertainty values (\texttt{nearestrefmagunc}) in quadrature. This gives the corrected errors in the measurements. 

\textit{Science Image Photometry:} We broadly follow the quality cut recommendations in the ZTF online documentation\footnote{\url{https://irsa.ipac.caltech.edu/data/ZTF/docs/releases/ztf_release_notes_latest}} and the ZTF Data System Explanatory Supplement.\footnote{\url{https://irsa.ipac.caltech.edu/data/ZTF/docs/ztf_explanatory_supplement.pdf}} We require that \texttt{catflags}$=$$0$. We also require that the PSF-fit \texttt{chi} and \texttt{sharp} parameters are within the same bounds as used in difference photometry. To be at par with the processing of the difference photometry, previously described, we reject all data points recorded under the secondary ZTF fields.

We call the data points passing all the above quality cuts as ``good.'' To avoid small-number statistics, we only consider objects having at least $40$ good data points in either $g$ or $r$-band in both standard and difference photometry. The final sample size following these steps is 490 objects. The median number of light curve data points is $\approx$$300$ (see Appendix \ref{app:stat}), sufficient for the subsequent analyses.

\subsection{Variability Metrics and Variable-Object Selection}\label{subsec:metrics_and_period}

\begin{deluxetable}{cc}
\tablenum{1}
\tablecaption{The selection and classification steps}
\label{tab:obj_num_table}
\tabletypesize{\small}
\tablewidth{0pt}
\tablehead{
    \colhead{Process} &
    \colhead{Number of Objects}
}
\startdata
\citetalias{Chornay21Distance} CSPNe with $>$50\% reliability                        & 1812              \\
ZTF data available               & 991              \\
40 ``good" ZTF detections in $g$- or $r$-band                       & 490               \\
Highly variable (HNEV) sources          & 94   \\
\hline
Short-timescale variables (this paper) & 83 \\
Long-timescale variables (Paper II) & 11 \\
\enddata
\tablecomments{For definition of HNEV sample, see Section \ref{subsec:metrics_and_period}.}
\end{deluxetable}

Variability metrics have been a powerful tool in identifying objects showing interesting light curve behaviors. Several of the commonly used metrics, however, suffer from systematic trends. Metrics like standard deviation, for example, show a positive correlation with magnitude just because of the increasing photon shot noise (see Appendix \ref{app:stat}). Either a detrending (for example, see \citealt{Guidry21}) or assignment of appropriate variability significance (see \citealt{Chornay21Binary}) is thus necessary. Unfortunately, with the limited starting sample, any such exercise is difficult here. Instead, we choose a metric that appropriately takes into account the photometric errors in the measurements, mitigating the effects of the systematics. We use the normalized excess variance (NEV) metric, similar to as defined in \cite{Coughlin21}:
\begin{equation}\label{eq:nev}
    {\rm NEV} = \frac{1}{N\widetilde{m}^2}\sum_i (m_i - \widetilde{m})^2 - \sigma_i^2
\end{equation}
where $N$ is the number of data points, $m_i$ are the magnitude values, $\widetilde{m}$ is the median and $\sigma_i$ are errors in the magnitude values\footnote{Unlike in \cite{Coughlin21}, we used the median, instead of mean, in the definition of NEV. This is because the median is resistant to outliers and thus a better measure of the ``quiescence" state of any object. We note, however, that, in this case, almost identical results are obtained if the mean is used instead.}. Any sample selection based on this metric is resistant to uncertain measurements. This is especially appropriate for our purpose where nebular contamination leads to higher photometric errors. Further discussion on the metric properties is presented in Appendix \ref{app:stat}.

\begin{figure}[t]
    \centering
    \includegraphics[width=\linewidth]{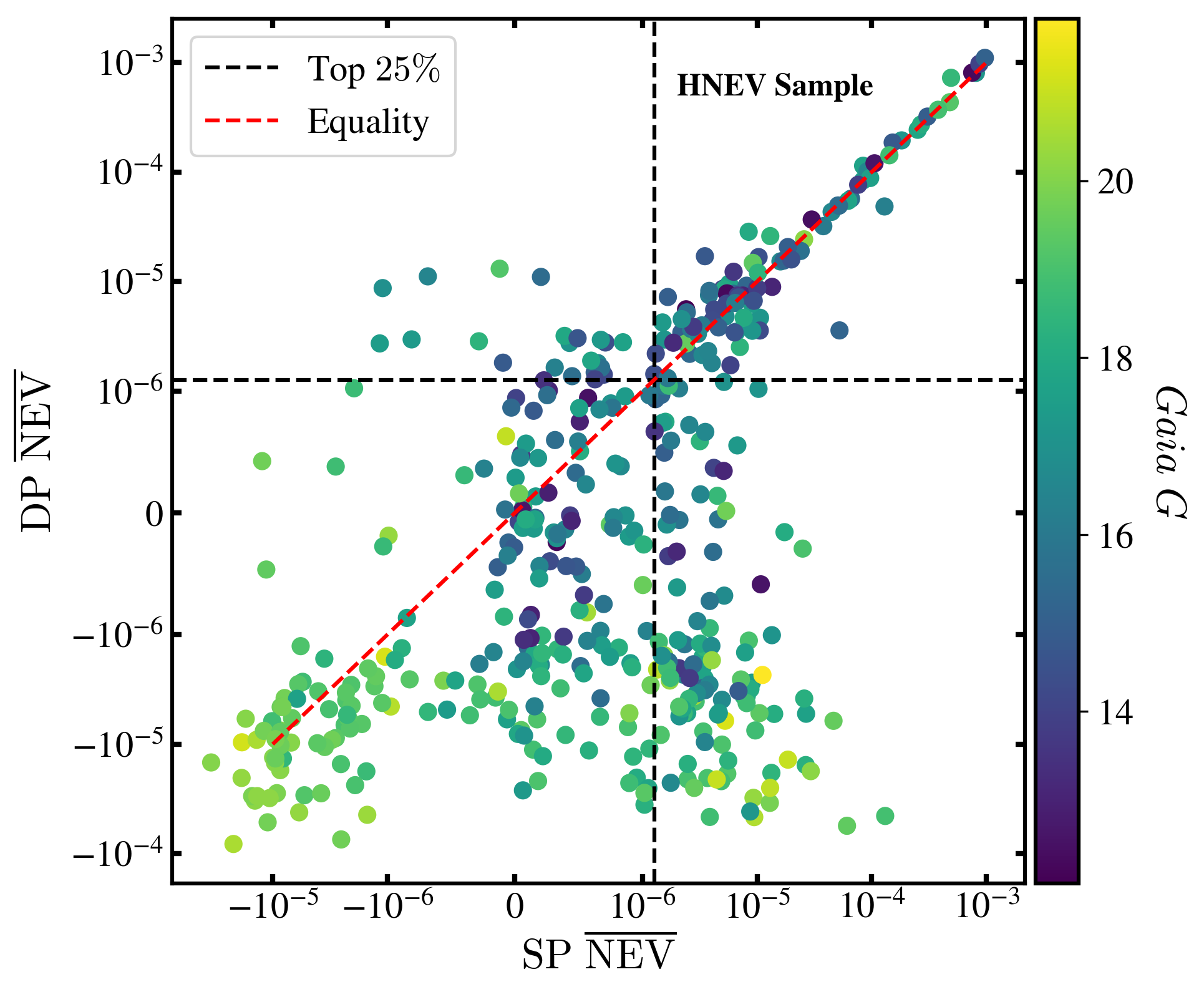}
    \caption{Position of all the CSPNe passing the ZTF photometric quality cuts in the space of the mean Normalized Excess Variance metric (${\rm \overline{NEV}}$, Equations \ref{eq:nev} and \ref{eq:mean_nev}) applied on the difference photometry (${\rm DP}$) and standard photometry (${\rm SP}$) light curves. The data points are colored according to \emph{Gaia} $G$-band magnitude. The horizontal dashed line marks the upper quartile in terms of ${\rm DP~\overline{NEV}}$ ($>1.25\times10^{-6}$). We define our variable (High-NEV, henceforth HNEV) sample where both ${\rm SP}$ and ${\rm DP~\overline{NEV}}$ exceed this threshold value (upper right quadrant, marked in the figure) to eliminate artifacts from either method of photometry.}
    \label{fig:nev_gaiag}
\end{figure}

We apply this metric separately to the $g$ and $r$-band light curves, on both the difference photometry (prefix ${\rm DP}$) and standard photometry (prefix ${\rm SP}$) light curves individually, after performing a 4$\sigma$ clipping of the data (to prevent severe outliers from affecting the metric). We do not apply the metrics to the $i$-band light curves which often do not meet the minimum data requirement. Following this, we define a band-averaged ${\rm NEV}$ metric as:
\begin{equation}\label{eq:mean_nev}
    {\rm \overline{NEV}} = \frac{n_g{\rm NEV_g}+n_r{\rm NEV_r}}{n_g+n_r}
\end{equation}
where $n_g$ and $n_r$ are the number of data points in $g$ and $r$ bands, respectively. Figure \ref{fig:nev_gaiag} shows the distribution of ${\rm DP~\overline{NEV}}$ and ${\rm SP~\overline{NEV}}$. The data points are colored according to \emph{Gaia} $G$-band magnitude. 

It is evident from the figure that ${\rm SP}$ and ${\rm DP~\overline{NEV}}$ agree quite well for the objects with high metric values (i.e. high variability). For objects without any significant intrinsic variability, the photometric errors ($\sigma_i$, see Equation \ref{eq:nev}) dominate the metric value. Due to certain systematics, the error values assigned to the standard photometric data are often underestimated, especially for fainter objects. This leads to disagreement between the ${\rm SP~\overline{NEV}}$ and ${\rm DP~\overline{NEV}}$, with the former often being larger than the latter. There are also a few objects showing low ${\rm SP~\overline{NEV}}$ but high ${\rm DP~\overline{NEV}}$ values. Manual inspection shows them to mostly be objects where artifacts from difference imaging (primarily the difference flux step discontinuity) are affecting the difference photometry light curve.

Despite the associated uncertainties, difference photometry-based ${\rm DP~\overline{NEV}}$ are better representative of the true variability, both due to cleaner light curves and more robust photometric errors. To define our highly variable sample, we use a metric threshold of $1.25\times10^{-6}$, which corresponds to the top 25\%\ in terms of ${\rm DP~\overline{NEV}}$. To minimize contamination from difference imaging artifacts, we require that the ${\rm SP~\overline{NEV}}$ is also greater than this threshold value. This forms our final selection of highly variable objects, which we call the High-NEV sample (HNEV, the upper right quadrant in the same figure). The size of this sample is 94. We restrict our further analyses to this subset of CSPNe. 

We briefly mention here a new avenue regarding the application of variability metrics to time series data. With the increasing amount of available data, a single metric often becomes insufficient to find hidden gems. This demands identifying suitable multi-dimensional metric spaces with easy interpretability and locating objects of interest within them. We introduce one specific two-dimensional metric space which, in the recent past, has shown significant promise with time series data. The two metrics are von~Neumann statistics \citep{VonNeumann41} and Pearson-Skew. Briefly, the former metric quantifies the orderliness in the variability (against random variability from photon noise), and the latter differentiates between dimming and brightening events. Though in this work we do not use this metric space for sample selection, we found that all the ``interesting'' variables discussed in our work have been well segregated in this space. We present this separately in Appendix \ref{sec:vonn_skewp} to share the discovery tool with the community. 

\section{Results}\label{sec:results}

\subsection{Categories of Variables}

We classify the variable central stars into two broad qualitative categories: short- and long-timescale variables. We call a variability ``short-timescale'' when the timescale of variations is shorter or comparable to the cadence of the ZTF survey (thus, the activity of the object is temporally not well resolved in the data). The number of such objects in the HNEV sample is 83, thus constituting the majority. If the light curve, on the other hand, shows variability\slash modulation\slash evolution over much longer timescales ($\gtrsim$1~month, for example) that is well resolved in ZTF, we call it a ``long-timescale'' variable. The remaining 11 objects fall in this category. In the rest of this paper, we focus on the short-timescale variables. A list of these variables is presented in Table~\ref{tab:short_timescale_vars}. The long-timescale variables will be discussed separately in Paper~II. Note that in all the figures that follow, we only show the difference-photometry--based light curves. Henceforth, we primarily work in the units of ``relative flux,'' which is obtained by normalizing the light curves to the median flux value. 

\subsection{Periodicity Search and Periodic Variables}\label{subsec:short_timescale_periodic}

\begin{figure*}[t]
    \centering
    \includegraphics[width=\linewidth]{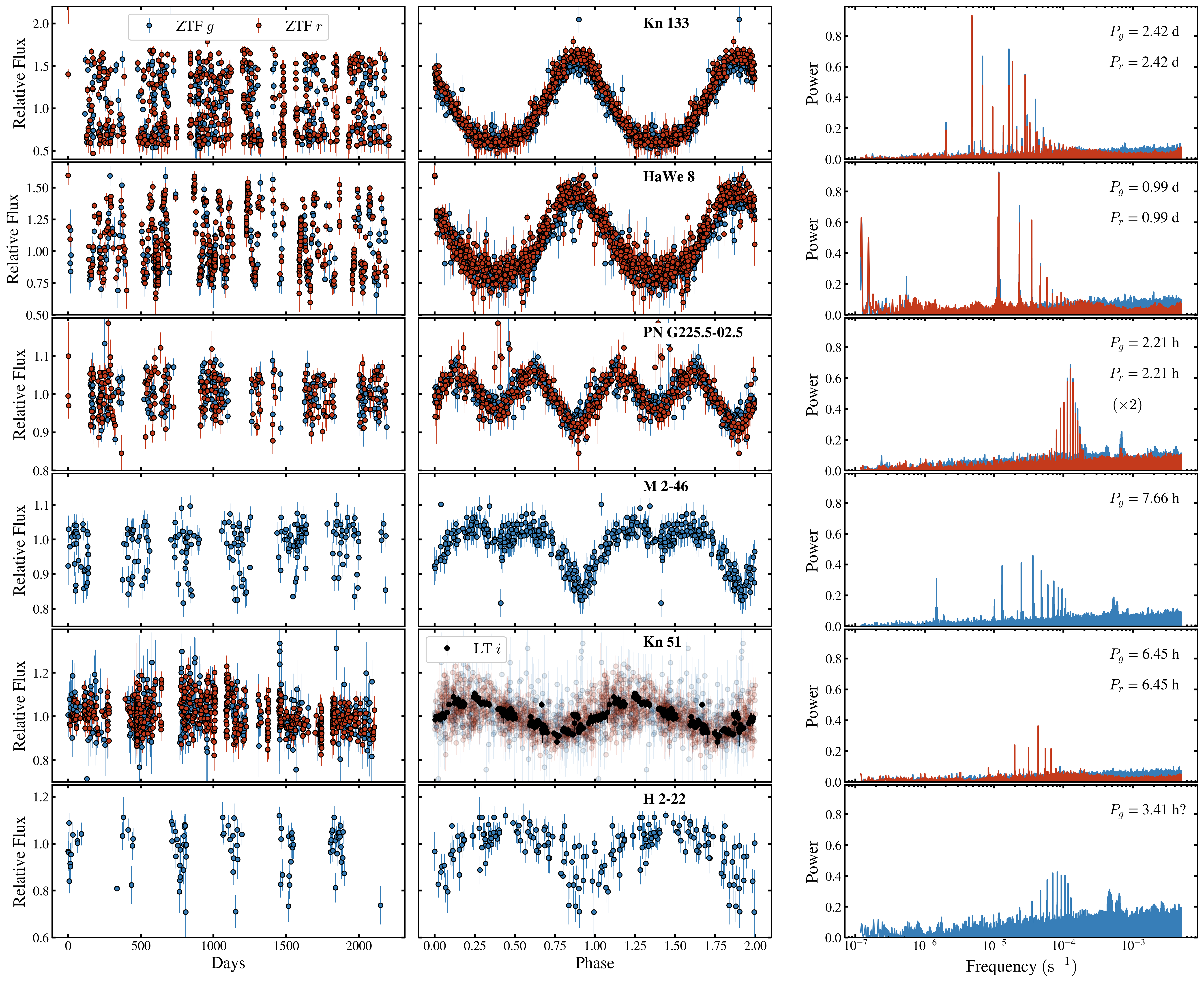}
    \caption{The newly discovered short-timescale periodic sources. \textbf{Left:} The raw ZTF light curve. \textbf{Center:} The ZTF (and LT for Kn~51) light curves phased at the best period. \textbf{Right:} The periodogram for both the $g$ and $r$ bands (nearly identical). We also quote the respective periods in the panels. The first five are binary candidates with confident period inference. H~2-22 is potentially periodic, although the inference is uncertain. The `$\times$2' in the right column indicates that the true period is likely twice the ZTF-inferred period (that is mentioned within the panels). In such cases, the twice of the period has been used to phase fold the light curves (center column). For a brief discussion on these objects see Section \ref{subsec:short_timescale_periodic}.}
    \label{fig:new_periodics}
\end{figure*}

We searched for significant periods in the light curves over a broad period range from $\sim$200 seconds (frequency of $5$~mHz) to $\sim$100~days ($0.12$~$\mu$Hz)\footnote{The lower limit of the period search is governed by the some of the shortest period binaries known (~few minutes, \citealt{Israe02,Burdge19}). The upper limit is chosen as a reasonable value for the short-timescale systems.} with a resolution of $5$~nHz. We first use the \texttt{astropy.timeseries} implementation of the Lomb-Scargle periodogram (\citealt{Lomb76,Scargle82,VanderPlas18}) to identify the most significant period. We then use this as the guess frequency to fit a one-component sinusoid to the data. This not only leads to a better identification of the period but also provides the amplitude and phase of the oscillation. Bearing in mind the possibility of different characteristics (period, amplitude, phase) of the light curves in the $g$ and $r$ bands, we performed a period search separately in the two bands. To establish consistency and potentially maximize the signal, we also perform a period search with the data combined from both bands (in units of relative flux). Following this, we manually reviewed the light curves and the periodograms. We considered both the strength of the periodogram peaks and also the morphology of the phase-folded light curves to determine if the period is trustworthy. Twenty-three objects showed a significant period.

We referred to the list of known binary candidates maintained by D. Jones\footnote{\url{https://www.drdjones.net/bcspn/}} and performed a SIMBAD\footnote{\url{https://simbad.cds.unistra.fr/simbad/}} search to identify the objects with known periods. We successfully recovered $17$ known periodic sources, with $15$ of them being binary candidates (see Figure \ref{fig:known_periodics_1}). Among them, the known period in WeDe~1 was detected only in the $r$-band data (with a subdominant peak at this period appearing in the $g$-band periodogram), and we failed to recover the period satisfactorily in Abell~30. In three sources among the remaining, (PM~1-23, PTB~26, and Hen~2-428), the ZTF-inferred period is half of the true period. In all the other objects, the correct periods were detected in both the ZTF bands individually. The two non-binary objects, PN~G162.1+47.9 \citep{Werner19} and Pa~27 \citep{Bond24} are rotational variables. We also recover an additional object, PN~M~3-2, where the variability arises from flux contamination from a bright unrelated binary which is $2''$ away from the CSPN \citep{Boffin18}. We recover a high-frequency alias of the actual period for the object, but we do not consider it among the periodic CSPNe. 

We report five new (to our knowledge) periodic variables in our HNEV sample with a confident period inference: Kn~133 (PN~G173.7-09.2), HaWe~8 (PN~G192.5+07.2), PN~G225.5-02.5, M~2-46 (PN~G024.8-02.7), and Kn~51 (PN~G164.8-09.8), and one candidate periodic sources with some uncertainty in the inference: H~2-22 (PN~G006.3+03.3). Figure \ref{fig:new_periodics} shows the raw and phase folded light curves along with the respective periodograms for the six sources. For two of the objects (M~2-46 and H~2-22), only the $g$-band light curve passed all the photometric quality cuts. For the other four objects, the periods inferred individually from the $g$ and $r$ bands and the combined data are consistent. Among them, follow-up $i$-band light curves with the Liverpool Telescope (LT) were available for Kn~133, HaWe~8, and Kn~51.\footnote{The observations were performed as a part of the PhD work of N. Chornay, see \cite{Chornay23}} An independent period search on this data confirms the periods of these objects. We present the results with LT data separately in Appendix \ref{appendix:follow_up_lt}. We note here that there are several more periodic variables in the full ZTF sample (see Appendix \ref{appendix:future_prospects} for three such examples). This complete study will be presented elsewhere (Tam et al. in prep). 


Four of these six objects, HaWe~8, M~2-46, Kn~51, and H~2-22 are currently labeled as ``True" PNe in HASH, which indicates that their nebular spectra predict their PN nature with high confidence. M~2-46 is also a known quadrupolar nebula \citep{Manchado95}. The remaining two objects, Kn~133 and PN~G225.5-02.5, are classified as ``Likely" PNe in HASH. An archival spectrum is present for PN~G225.5-02.5 on the HASH website, which bears a close resemblance to typical PNe spectra. All of the objects have resolved nebular counterpart. 

The light curves of both Kn~133 and HaWe~8 bear the characteristic shape of an irradiation effect (see also Appendix \ref{appendix:follow_up_lt} for the LT light curves). For PN~G225.5-02.5, inspection showed that the ZTF-inferred period is half of the true period of $4.42$~hours. The light curve resembles ellipsoidal variability and bears resemblance to other known binaries like PM~1-23 and Hen~2-428 (see Figure \ref{fig:known_periodics_1}). M~2-46 appears to be an eclipsing system, with a deep primary eclipse and a shallow secondary eclipse. The shape of the light curve bears resemblance to previous known systems (like PHR~J1725-2338, see \citealt{Jacoby21}), which indicates that it is a combination of both eclipse and ellipsoidal modulation. 

For Kn~51, a previous indication of periodicity in the literature appears in \cite{Chornay22}, where they mention a period detection in \emph{Gaia} photometric data. The period, although, was not reported. We detect a significant period of 6.45~hours in ZTF, consistently in both the bands. The period was confirmed with the LT $i$-band light curve. The ZTF data for this object is noisy with large photometric errors. Thus, we present the phase-folded LT data in Figure \ref{fig:new_periodics} for clarity. For the purpose of this figure, we use the ZTF period (which gives a more precise measurement owing to its much longer baseline) to phase fold the LT data. See Appendix \ref{appendix:follow_up_lt} for the independent results from the LT data.

The raw ZTF light curve for H~2-22 shows consistent dips of similar amplitudes throughout the ZTF baseline. A significant period of $3.41$~hours was found. The phase-folded light curve resembles ellipsoidal variability and/or eclipse (in fact, similar to the primary eclipse in M~2-46). However, the significant scatter, especially near the minimum, indicates uncertainty in the inferred period. The inferred period can also be an alias of the true period, especially with a doubly eclipsing system. Different algorithms also lead to slightly different period inferences, adding to the uncertainty. For example, a phase dispersion minimum analysis gives a period of 3.97~hours, though the phase folded light curve looks worse. Thus, we do not confidently claim the periodicity of this object. We note here that short-timescale photometric modulation (in under 2 hours) in this object was also detected in \cite{Toma16}.

\subsection{Sources without significant period: ``Aperiodic''}\label{subsec:short_timescale_aperiodic}

\begin{figure}[t]
    \centering
    \includegraphics[width=\linewidth]{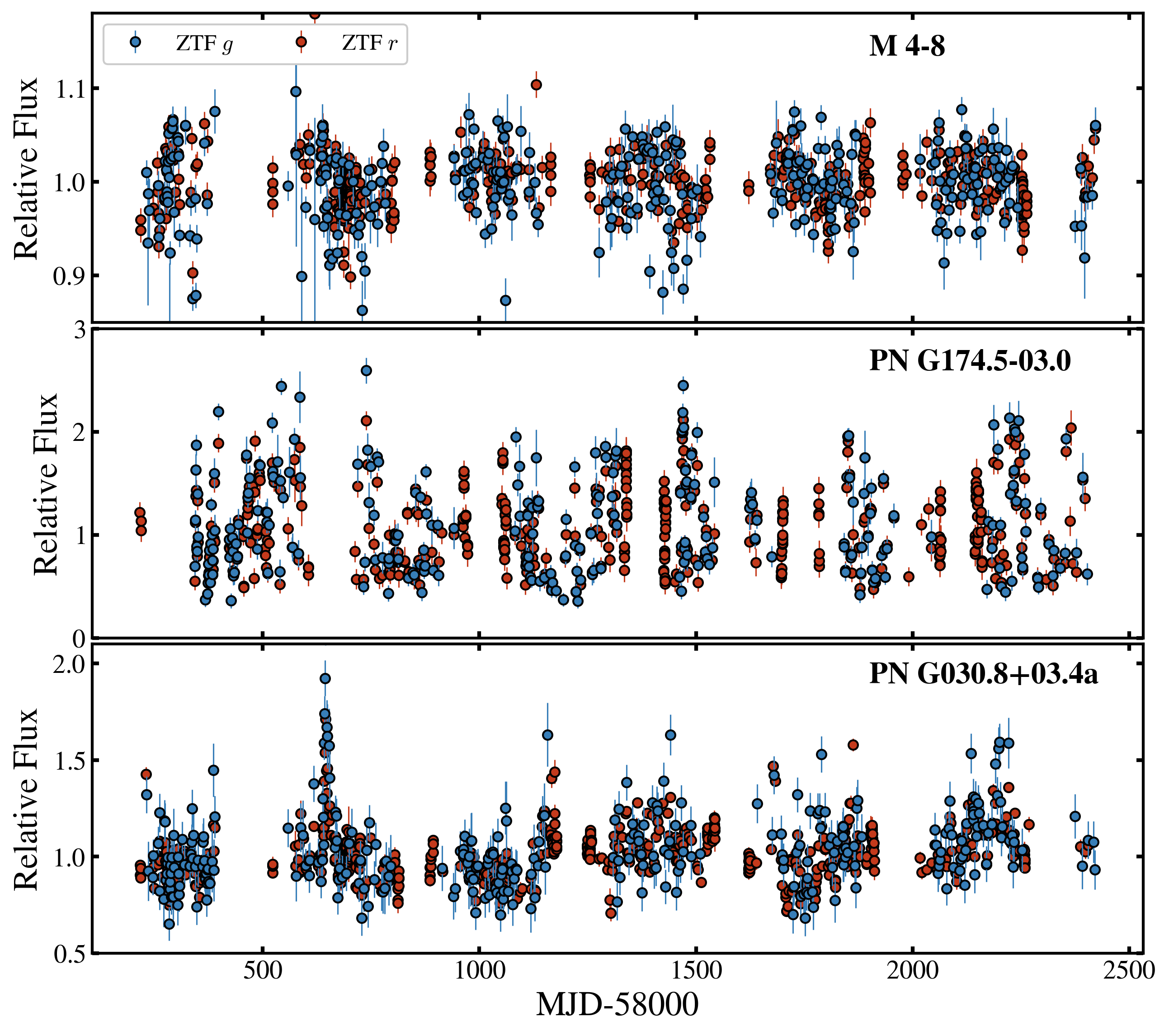}
    \caption{Examples of light curves for two types of short-timescale aperiodic variables. M~4-8 shows variability typical of those induced by stochastic wind activity from hot central stars. Both the others are outbursting objects. Spectroscopic observation of PN~G030.8+03.4, however, makes it less likely to be a PN (see Appendix \ref{appendix:png030spectrum}). For a brief discussion on this class of objects, see section \ref{subsec:short_timescale_aperiodic}.}
    \label{fig:new_short_vars}
\end{figure}

The remaining 63 objects in our HNEV sample do not show a significant period in the ZTF light curves. We term all of them as ``aperiodic variables.'' However, we cannot rule out the possibility of a few periodic sources, with periods undetected in ZTF, to be misclassified to this category (a more careful and detailed periodicity search will be presented in Tam et al., in prep). Among these sources, the majority show consistent variability about a median flux with peak-to-peak flux amplitude between $\sim$10 -- 40\% (we call them jittering sources). We conducted a thorough search on SIMBAD to look for any known source of variability to any of the objects. Variability in four of them: NGC~1501 (PN~G144.1+06.1), LW~Lib (central star of Fr~1-5 or PN~G337.9+43.5), Hen 2-260 (PN~G008.2+06.8), and PM~1-308 (PN~G034.5-11.7) has been attributed to winds and/or pulsations of the CSPN. Note that a 3.3~day period in NGC~1501 was recently detected in \citet{Aller24}, which we did not detect in ZTF. This is, thus, an example of misclassification and also indicates a possible failure of simple Lomb-Scargle to identify periodicity with complex light curve morphology (see Figure A.1 in their paper). One of the rest, GK~Per (PN~G150.9-10.1) is a famous system of classical nova inside an old planetary nebula \citep{Bode87}. The system has an orbital period of $\sim$$2$~days \citep{Crampton86,Alvarez21}. We do not detect the period in ZTF primarily due to outbursts and nova activity contaminating the light curve over the long baseline. We tried to be as complete as possible in our search, though we do not rule out the possibility of missing out on a few known variables among those remaining in the list. 

Figure \ref{fig:new_short_vars} shows example light curves of two types of aperiodic short-timescale variables identified in this study. The top panel, showing M~4-8, is a typical example of a jittering source. The archival spectrum on HASH indicates it to be a very low excitation (VLE) young and compact PN. Thus, the likely source of variability is wind activity and\slash or unresolved pulsations from the young CSPN. A few objects were also seen to show outbursts. Two prominent examples are shown in the bottom two panels of the same figure. PN~G174.5-03.0 displays regular high-amplitude outbursts, whereas PN~G030.8+03.4a (not to be confused with PN~G030.8+03.4 a.k.a Abell 47) shows a single outburst followed by shorter amplitude variability. Such behaviors are potentially interesting owing to the possibility of accretion\slash mass-transfer events inside a PN. However, these sources are labeled `Possible PN' on HASH without a spectroscopic observation and, thus, may undergo reclassification. For example, the recently acquired spectrum for PN~G030.8+03.4a makes it more likely to be a symbiotic system or a cataclysmic variable over a PN (see Appendix \ref{appendix:png030spectrum}). Interestingly, this object appears as a candidate wide-binary system in \cite{Gonzalez21Edr3}, with an M-type companion at a distance of 5000~AU from the central star. If it is an interacting system, it has to have a closer companion, making it a candidate for a hierarchical triple.

\begin{figure*}[t]
    \centering
    \includegraphics[width=\linewidth]{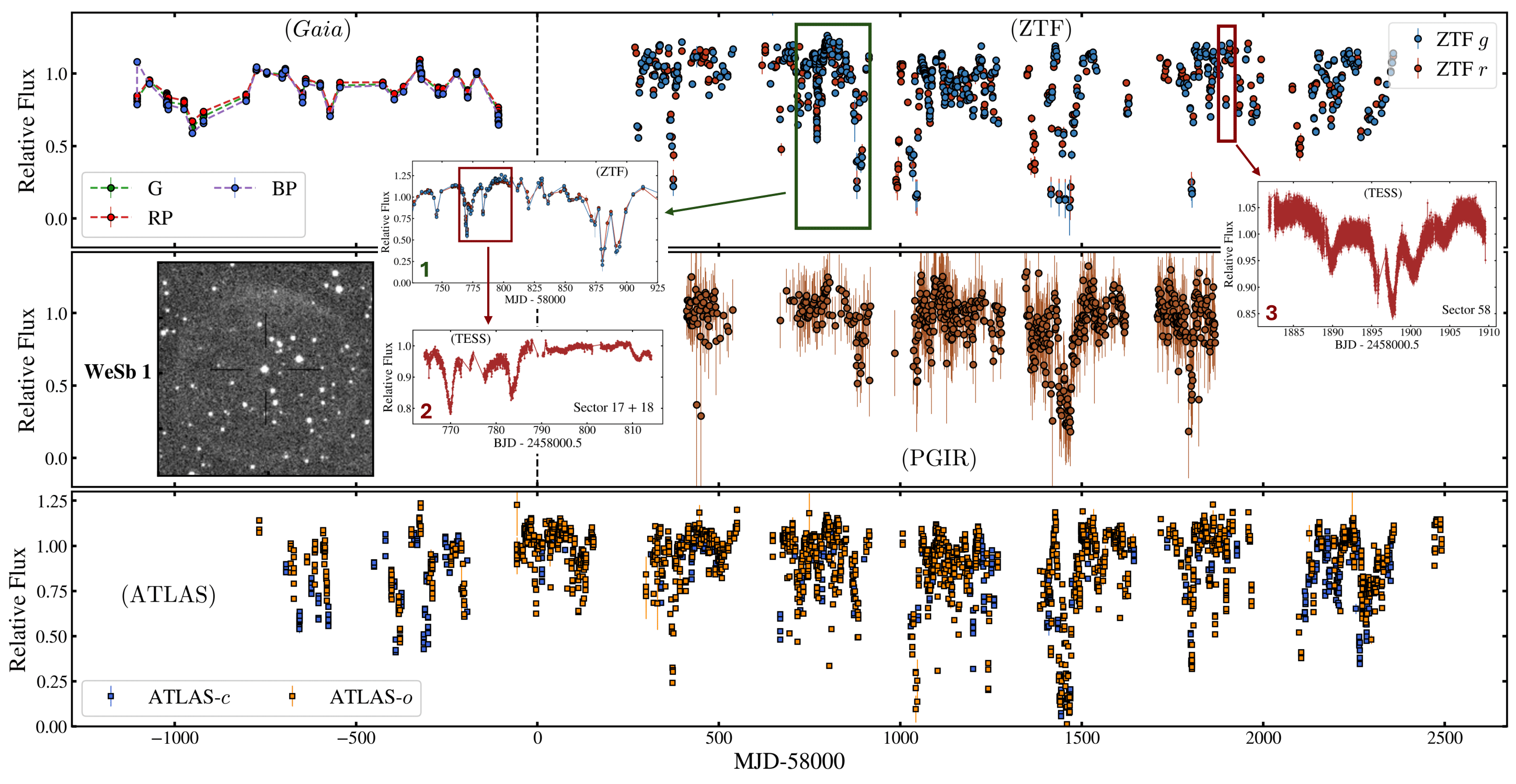}
    \caption{The light curves of WeSb~1 from different instruments: \emph{Gaia} (top left), ZTF (top right), Palomar Gattini-IR (PGIR, middle right), ATLAS (bottom), and TESS (as insets). Irregular deep (often $\sim$100\% in flux) dips are evident in ZTF, ATLAS, and PGIR light curves. Significant variability is also seen in \emph{Gaia} light curve. The three numbered insets are: 1) Zoom-in of a portion of the ZTF light curve where the activity is temporally well resolved, 2) the TESS light curve of a sub-portion, and 3) TESS light curve of a different portion of the light curve. The transit-like features are well-resolved in the TESS light curves, though the amplitudes are underestimated (due to contamination from nearby stars within the large TESS PSF). We also provide the narrow-band ${\rm H\alpha}$ image (middle left) from the Palomar Transient Factory (PTF) Census of Local Universe (CLU) survey \citep{Law09, Cook19}, with the nucleus marked. The large and evolved ring-like nebula is visible. See Section \ref{appendix:wesb1} for further discussion on this object.}
    \label{fig:wesb1_ztf_gattini_ir}
\end{figure*}

One object, WeSb~1 (PN~G124.3-07.7), does not belong to the above categories and defines its own variability class: aperiodic dipper. The light curve shows prominent, persistent, short-lived deep transit-like dips, often obscuring $\sim$100\% of the flux from the source (see Figure \ref{fig:wesb1_ztf_gattini_ir}). Without any significant period, it appears unlikely to be a typical eclipsing binary. Rather, the complex morphology of the transits resembles those incurred by debris discs (for example, white dwarfs with transiting planetary debris; see Figure 6 in \citealt{Guidry21}). This motivated further studies, and we present our analyses on this object separately in Section \ref{appendix:wesb1}. 

\subsection{Comparison with previous works}

Several of our HNEV objects appear in the list of \emph{Gaia} variables of \cite{Chornay21Binary}. Out of their 57 objects (Table A.1 in their paper), ZTF light curves could be obtained for 29 objects, the rest of them being situated outside the ZTF footprint. Only 21 of these passed the imposed photometric quality cuts. Out of them, we could successfully recover 20 of them in our HNEV sample, 19 of which are short-timescale variables. Five of them are periodic variables (Cr~1, KTC~1: periods inferred in \citealt{Chornay21Binary} and this work; Pa~27: \cite{Bond24} and this work; Kn~133 and HaWe~8: this work), and the rest are aperiodic variables which include WeSb~1. This high recovery rate demonstrates the success of the variability metric with ZTF photometry. Additionally, this also gives confidence in the variable nature of these CSPNe, being identified as such independently in two surveys with vastly different analysis methodologies.

We also cross matched our sample with the known short-period binaries in the website by D. Jones. ZTF data was acquired for 47 objects, out of which only 25 passed the quality cuts. Among them, 17 objects were recovered in the HNEV sample. The periods were recovered in 15 of them (see \ref{fig:known_periodics_1}) except for NGC~1501 and GK~Persei. Recently, \cite{Ali23} conducted a search for variable CSPNe using the \texttt{phot\_variable\_flag} identifier in Gaia DR3 data. Through the CSPN list of \citetalias{Chornay21Distance}, ZTF data for 30 of their variable candidates could be queried, out of which 18 passed the photometric quality cuts. Among them, 12 objects have been recovered in our HNEV sample which includes several known binaries.

Though the recovery rate of known variables is high, it is not complete. This shows that several more variable\slash binary sources remains to be discovered beyond the HNEV sample. We substantiate this with discovery of three new periodic sources outside of the HNEV sample in Appendix \ref{appendix:future_prospects}.

\section{W\MakeLowercase{e}S\MakeLowercase{b}~1}\label{appendix:wesb1}

WeSb~1 is a ``Likely'' PN in HASH. Its extended nebula appears as a ring (see the narrow-band ${\rm H\alpha}$ image in the middle-left panel in Figure \ref{fig:wesb1_ztf_gattini_ir}) with an angular diameter of 185$''$. It was first detected by interference-filter photography by \cite{Weinberger81}. \cite{Rosado91} performed deep narrow-band photography, detecting the ring-like nebula as well as emission from its interior (also see deep ${\rm H\alpha}$ amateur image in \url{https://www.astrobin.com/vgr1j3/?q=wesb%201} revealing the nebula in its entirety). \cite{Pereyra13} considers WeSb~1 to be a highly evolved PN. The inferred distance to the central source of WeSb~1 (henceforth referred to as the nucleus) from \emph{Gaia} parallax is $3.7\pm0.3$~kpc \citep{Gaiadr3}. A similar distance of $3.3\pm0.3$~kpc was obtained in \citet{Bailer-Jones}, where three-dimensional Galactic models were used to construct priors to improve the distances inferred from \emph{Gaia} data. This makes the physical diameter of the ring to be $\sim$3~pc. We note here that a compatible distance and diameter ($2.4\pm0.7$~kpc and 2.2~pc, respectively) to the nebula was obtained by \cite{Frew16} using the correlation between PN size and ${\rm H\alpha}$ surface brightness. The agreement between the two estimates (within 2$\sigma$) provides confidence in the correct identification of the central stellar source. \citet[][another work using \emph{Gaia} eDR3 to assign central stars to PNe]{Gonzalez21Edr3} also identifies this source to be the nucleus of WeSb~1, albeit with a quality flag of `B' (due to the \emph{Gaia} color not being sufficiently blue, the possible reason for which will be clear in the following discussion). We also performed a brief check (presented in Appendix \ref{appendix:misidentified}) and do not find any obvious evidence against the current source being the true nucleus of WeSb~1. Thus, we proceed with our analyses. Henceforth, we assume the \emph{Gaia} distance of $3.7$~kpc for the nucleus of WeSb~1. Assuming any other distance within the bounds of the \emph{Gaia}-based inferences does not alter our conclusions significantly. The summary of the analyses is presented in Table \ref{tab:wesb1_table}.

\begin{deluxetable}{lcr}
\tablenum{2}
\tablecaption{The summary of the candidate binary nucleus of WeSb~1. See Section \ref{appendix:wesb1} for the details of the analyses.}
\label{tab:wesb1_table}
\tabletypesize{\small}
\tablewidth{0pt}
\tablehead{
    \colhead{Parameter} &
    \colhead{} &
    \colhead{Value}
}
\startdata
&\emph{Gaia} Properties&\\
DR3 ID && 423384961080344960 \\
RA (deg) && 15.22543 \\
Dec (deg) && 55.06667 \\
Parallax (mas)&&0.271$\pm$0.025 \\
$G$ && 14.772$\pm$0.006 \\
$G_{\rm BP}-G_{\rm RP}$&&1.17$\pm$0.04 \\
$\mu_{\alpha}$ (mas~yr$^{-1}$)&&-1.28$\pm$0.02 \\
$\mu_{\delta}$ (mas~yr$^{-1}$)&&-1.29$\pm$0.02 \\
Radial Velocity && -- \\
\hline
&Inferred Properties&\\
Electron Density && $\gtrsim$$10^6~{\rm cm^{-3}}$\\
Companion $T_{\rm eff}$&&$\sim$6500~K\\
Companion radius&&$\sim$6~$R_{\odot}$\\
Companion mass&&$\sim$$2.3$~$M_{\odot}$\\
Companion spectral type&&F-type subgiant?\\
Companion Age && $\sim$$0.75$~Gyr\\
CSPN $T_{\rm eff}$&&$>$$10^5$~K\\
CSPN radius&&$>$$0.05~R_{\odot}$\\
CSPN initial mass and age && Same as companion? \\
\enddata
\tablecomments{The values corresponding to the inferred properties are tentative, mostly based on visual comparison of models and data. Thus we do not quote formal errors. See the text for the details.}
\end{deluxetable}

\subsection{Light curves}\label{subsec:wesb1_lcs}

The top right panel in Figure \ref{fig:wesb1_ztf_gattini_ir} shows the ZTF light curve of WeSb~1, where very deep and irregular dipping features are evident. Portions of the light curve are well-resolved, where the eclipses can be seen in and out. A zoom-in of one such section is presented as an inset in the same figure. The ingress and egress appear to occur over a few days, but the minima are short-lived. As a supplement, we also provide the optical light curves from the Asteroid Terrestrial-impact Last Alarm System (ATLAS) survey (bottom panel in the same figure) as it has a longer baseline than ZTF\footnote{We queried both the difference and science-image forced photometry data from the online service at \url{https://fallingstar-data.com/forcedphot/} \citep{Shingles21}. Owing to their warnings about the latter, we primarily use the former but use the science-image data for estimation of the reference flux and normalization. We reject all data points where \texttt{err}$\neq$0 and the photometric errors are $\geq$$20$\%.}. It primarily operates in two bands: cyan (4200 -- 6500~\AA) and orange (5600 -- 8200~\AA). The variability is recovered in ATLAS data, showing dips of similar depth coincident with those in ZTF. Epochal photometric data from \emph{Gaia} is also available for this source (top-left panel in the figure). The variability amplitude here is $\sim50\%$, much smaller than in ZTF\slash ATLAS. However, this can just be due to the insufficient cadence of \emph{Gaia}, failing to resolve the deep eclipses occurring on shorter timescales. 

We extracted light curves from the Transiting Exoplanet Survey Satellite (TESS) mission for the nucleus of WeSb~1 using the online tool TESSExtractor\footnote{\url{https://www.tessextractor.app/}} \citep{Brasseur19,Serna21}. The light curves are provided as insets in the same figure. We note here that the data from Sectors 17 and 18 (inset 2 in the figure) is heavily affected by TESS systematics and should be read with caution. The data from Sector 58 (inset 3) is much less affected and, thus, more reliable. The transit-like dips are prominent and well resolved in the TESS light curves. The amplitudes, however, appear to be underestimated, possibly due to the contamination from nearby sources due to the large PSF of TESS. No obvious periodicity is evident in the data.

To investigate the behavior of the source in other wavelengths, we present light curves from Palomar Gattini-IR (PGIR) observations, operating in the near-infrared $J$ band\footnote{PGIR is a wide-field near-infrared time domain survey covering the entire accessible sky with a cadence of $\approx 2$\,nights. We used the publicly available catalog of forced PSF photometry light curves for all 2MASS sources in the PGIR observing footprint, as described in \citet{Murakawa2024}.}. Though the associated errors are large due to the lower sensitivity of PGIR, the major activities are recovered in the light curve. 

\begin{figure}[t]
    \centering
    \includegraphics[width=\linewidth]{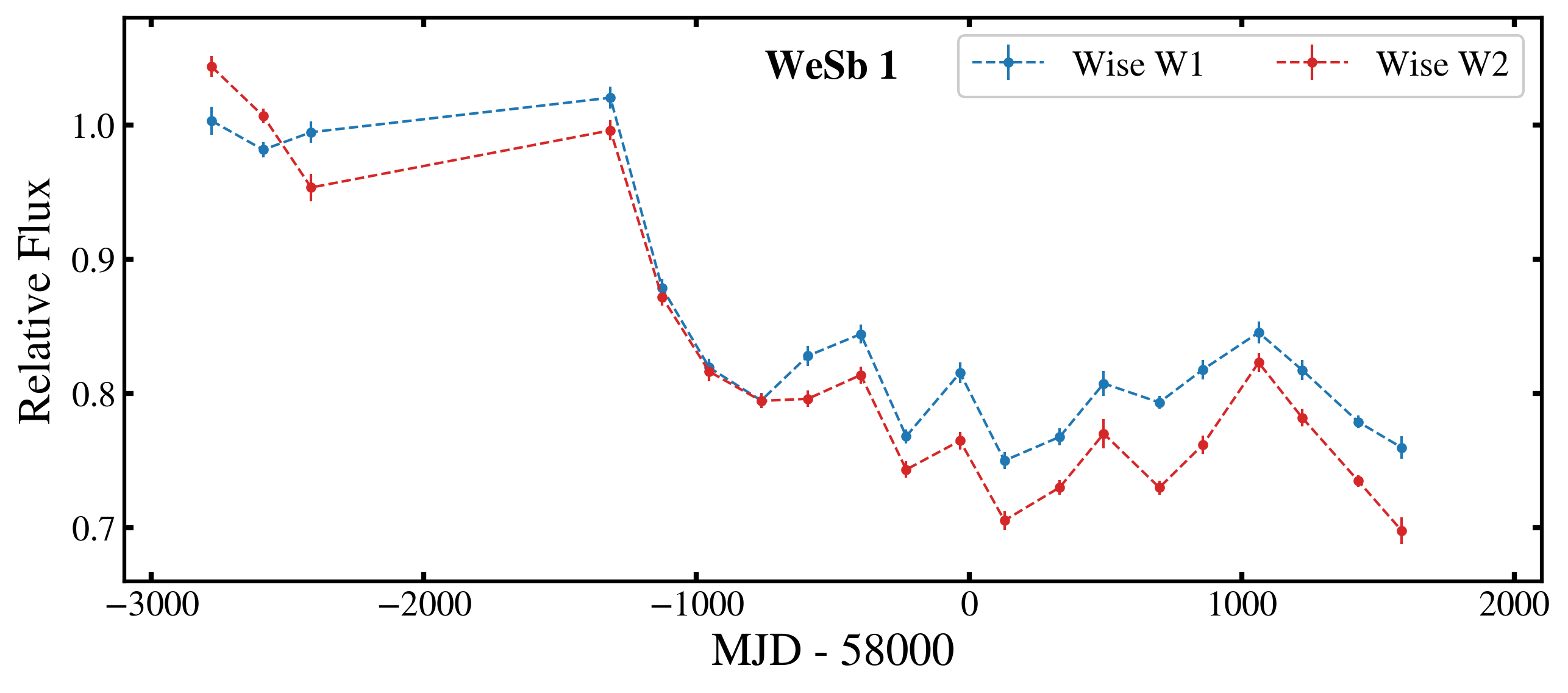}
    \caption{The WISE lightcurve for WeSb~1. Significant variability (albeit of lower amplitude) is seen over the baseline of the light curves from the other surveys presented in Figure \ref{fig:wesb1_ztf_gattini_ir}. This is preceded by a sharp drop in flux around ${\rm MJD-58000\sim-1200}$. Though not evident in this figure, a gradual blueing by $\sim$$0.1$ magnitude is present over the whole baseline.}
    \label{fig:wesb1_wise}
\end{figure}

Within the error bars, the amplitudes of the dips agree in all the photometric bands from ZTF through PGIR (see for example the major event at ${\rm MJD-58000\sim1500}$). This is suggestive of achromatic variability, which enables us to put some constraints on the grain size of the obscuring material. In Mie scattering, the scattering cross section becomes wavelength independent for sufficiently large size parameters, ${X = 2\pi a/\lambda}$, where $\lambda$ is the wavelength, and ${a}$ is the grain radius. Studies with regards to debris discs in other systems (see for example \citealt{Croll14}, \citealt{Xu18}) shows that $X$$\gtrsim$$2$ is sufficient for achromaticity. The assumption that the dips are achromatic up to the ${J}$ band (centered on ${1.25~\mu \rm m}$), puts a lower bound on the particle size of ${\rm 0.4~\mu m}$. This is no surprise as in most systems with dust/debris, the grain sizes are usually larger. 

We also provide the epochal forced-photometry data from the Wide-field Infrared Survey Telescope (WISE, \citealt{Wright10}; figure \ref{fig:wesb1_wise})\footnote{Forced aperture photometry was performed at the source position in publicly available time-resolved \texttt{unwise} images \citep{Meisner2018}, as described in \citet{Tran2024}.}. Significant variability is evident. The intriguing feature is the sharp drop in luminosity in both the WISE bands at around ${\rm MJD-58000\sim-1200}$. This event precedes any of the other surveys, thus not seen in their light curves. Jittering variability is observed over the temporal baseline of \emph{Gaia} and ZTF, but the amplitude is significantly smaller ($\sim$15\%). As with \emph{Gaia}, this may also be a result of the insufficiency of the WISE cadence to resolve the shorter-timescale activities. Though not obvious from the figure, mild and gradual bluing by $\sim$$0.1$~magnitude in the WISE bands (brightening of W1 relative to W2) is observed from the beginning to the end of the WISE coverage.

\subsection{Optical Spectroscopy}\label{wesb1_spectrum}

\begin{figure*}[t]
    \centering
    \includegraphics[width=\linewidth]{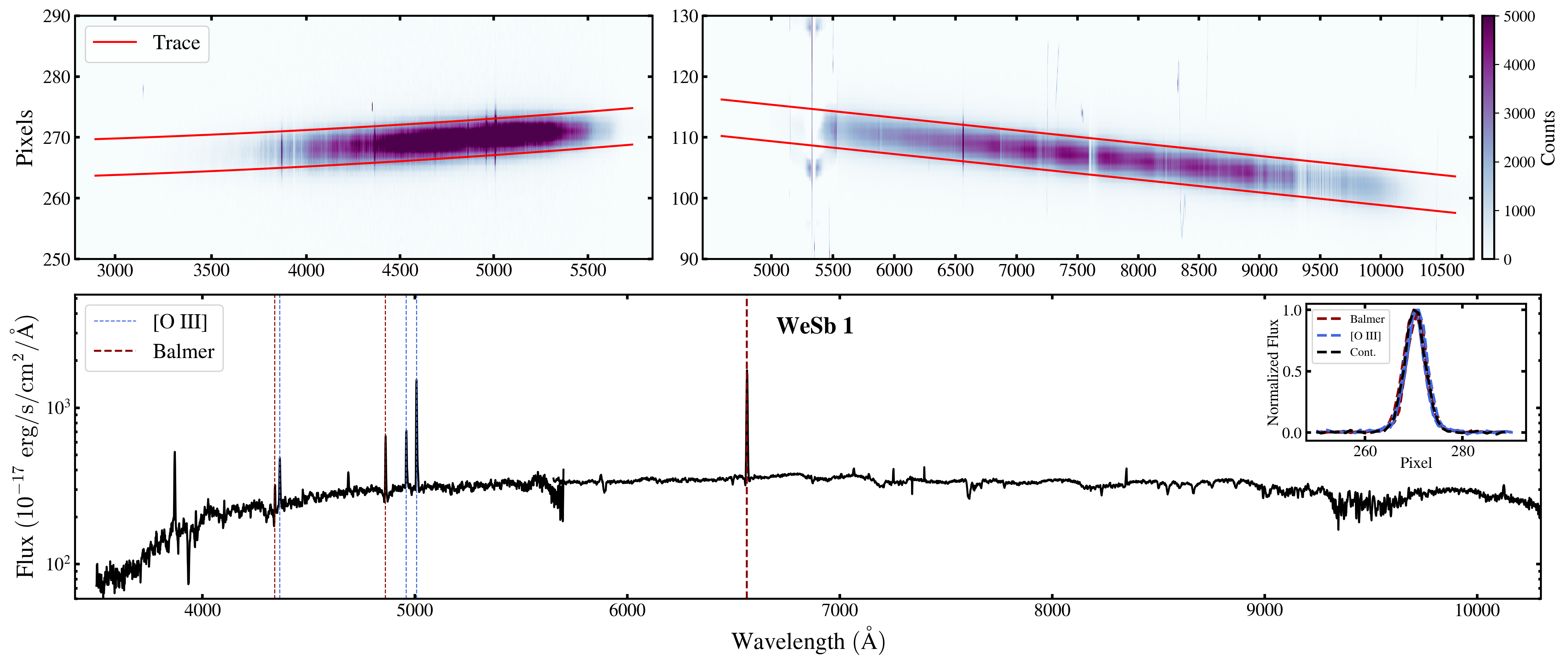}
    \caption{The background-subtracted DBSP spectrum for WeSb~1. The upper left and right panels show the two-dimensional slit spectra for the blue and the red arms respectively. The bottom panel shows the one-dimensional spectrum derived from the marked traces. We also mark the first three Balmer and [\ion{O}{3}]~4364,~4959,~5007~\AA\ emission lines in the spectrum. The spatial profiles (i.e. along the vertical axis in the two-dimensional spectra) of these emission lines along with that of the continuum are presented as an inset in the bottom panel. This agreement shows that the emission lines are a part of the nucleus.}
    \label{fig:WeSb1_trace_spectrum}
\end{figure*}

\subsubsection{Observations}\label{subsubsec:wesb1_obs}

\textit{Palomar\slash DBSP}: We obtained optical spectroscopic data of WeSb~1 using the DouBle SPectrograph (DBSP; \citealt{Oke82}) attached to the Cassegrain focus of the Palomar 200-in Hale Telescope. The observation was performed on 2024 July 12, with a 10-minute exposure. We used the D55 dichroic, the 600 line~mm$^{-1}$ grating for the blue arm blazed at 3780~\AA, and the 316 line~mm$^{-1}$ grating for the red arm blazed at 7150~\AA. We used grating angles of 27$^{\circ}$17$'$ and 24$^{\circ}$38$'$ for the blue and red sides, respectively. With these setups and a slit width of 1.5$''$ long-slit, we achieve resolving powers of $R=1600$ in the blue arm and $R=1400$ in the red arm. These setups provide continuous spectral coverage across 2900--10800~\AA, with the division between blue and red arms typically occurring around 5650~\AA. The standard star BD\,+33$^\circ$2642 was used for flux calibration. The seeing was $\sim$$1.5-2''$ when the observation was performed.

The standard reduction pipeline for DBSP spectra, DBSP\_DRP \citep{Roberson21}, based on PypeIt \citep{PypeIt_1, PypeIt_2} was found unsuitable for PNe, which often have strong emission lines. The optimal reduction procedure of DBSP\_DRP was often masking the bright and extended emission lines. Thus, we manually run PypeIt to perform the reduction. We follow the PypeIt recommendations for bright emission lines, and set the parameters \texttt{use\_2dmodel\_mask = False} and \texttt{no\_local\_sky = True}, to avoid masking and local subtraction of the lines. Following this, we perform manual trace identification on the two-dimensional spectrum. We use a trace width of six pixels in the spatial direction, centered on the stellar trace to obtain the one-dimensional spectrum. We note here that, for a yet unidentified issue, the sensitivity function for the red arm could not be generated for the night with BD\,+33$^\circ$2642 as the standard star. Thus, we used the sensitivity function from a previous observing run (May 13, the observation date for the long-timescale variables to be described in Paper II) generated using Feige\,34. A step discontinuity between the red and the blue arms was witnessed in the resultant one-dimensional spectrum of WeSb~1. The red arm spectrum was manually scaled to match the continuum of the blue arm. Finally, we compared the spectrum to the spectral energy distribution (SED) of the source (presented in section \ref{subsubsec:wesb1_sed}). Possibly due to suboptimal calibration and/or slit losses, the DBSP flux was a factor of $1.6$ lower. This can also arise from the variability of the object, with the object being dimmer at the epoch of observation. We upscaled the spectrum to match the SED.

\textit{HET\slash LRS2-B}: The central star of WeSb~1 was also observed with the Low-Resolution Spectrograph (LRS2) on the 10-m Hobby-Eberly Telescope (HET; \citealt{Ramsey98, Hill21}) located at the McDonald Observatory in West Texas, USA, as a part of a survey for faint PN central stars (undertaken by H.E.B, see \citealt{Bond23} and subsequent papers) on 2024 August 9 and 26, both with a 3-minute exposure.  The LRS2 instrument is described in detail by \cite{Chonis14,Chonis16}. Briefly, LRS2 provides integral-field-unit (IFU) spectroscopy with 280 $0\farcs6$-diameter lenslets that cover a $12'' \times 6''$ field of view (FOV) on the sky. LRS2 is composed of two arms: blue (LRS2-B) and red (LRS2-R). All of our observations are made with the target placed in the LRS2-B FOV. The LRS2-B arm employs a dichroic beamsplitter to send light simultaneously into two spectrograph units: the ``UV'' channel (covering 3640--4645 \AA\ with a resolving power of 1910), and the ``Orange'' channel (covering 4635--6950 \AA\ with a resolving power of 1140). Data reduction was performed using the pipelines developed by G.R.Z.\footnote{\url{https://github.com/grzeimann/Panacea}}$^{\rm ,}$\footnote{\url{https://github.com/grzeimann/LRS2Multi}} (Zeimann et al., in prep.). We obtain both narrow-band images around primary emission lines and a one-dimensional spectrum extracted from the fibers containing the WeSb~1 nucleus. Similar to DBSP, the one-dimensional spectrum was scaled up (by a factor of $1.5$) to match the continuum with the SED.

Figure \ref{fig:WeSb1_trace_spectrum} shows the background-subtracted spectra for the central star of WeSb~1 obtained using DBSP. The HET spectrum has been provided in Appendix \ref{appendix:wesb1_het} (top panel of Figure \ref{fig:WeSb1_spec_compare_HET}), and is in reasonable agreement with the DBSP spectrum. Due to systematics associated with both observations, the quantitative estimates (like inferred emission-line fluxes and equivalent widths of absorption\slash emission lines) from the two spectra are slightly different. We use this to estimate errors in the measurements presented in the following section.

\begin{figure}[t]
    \centering
    \includegraphics[width=\linewidth]{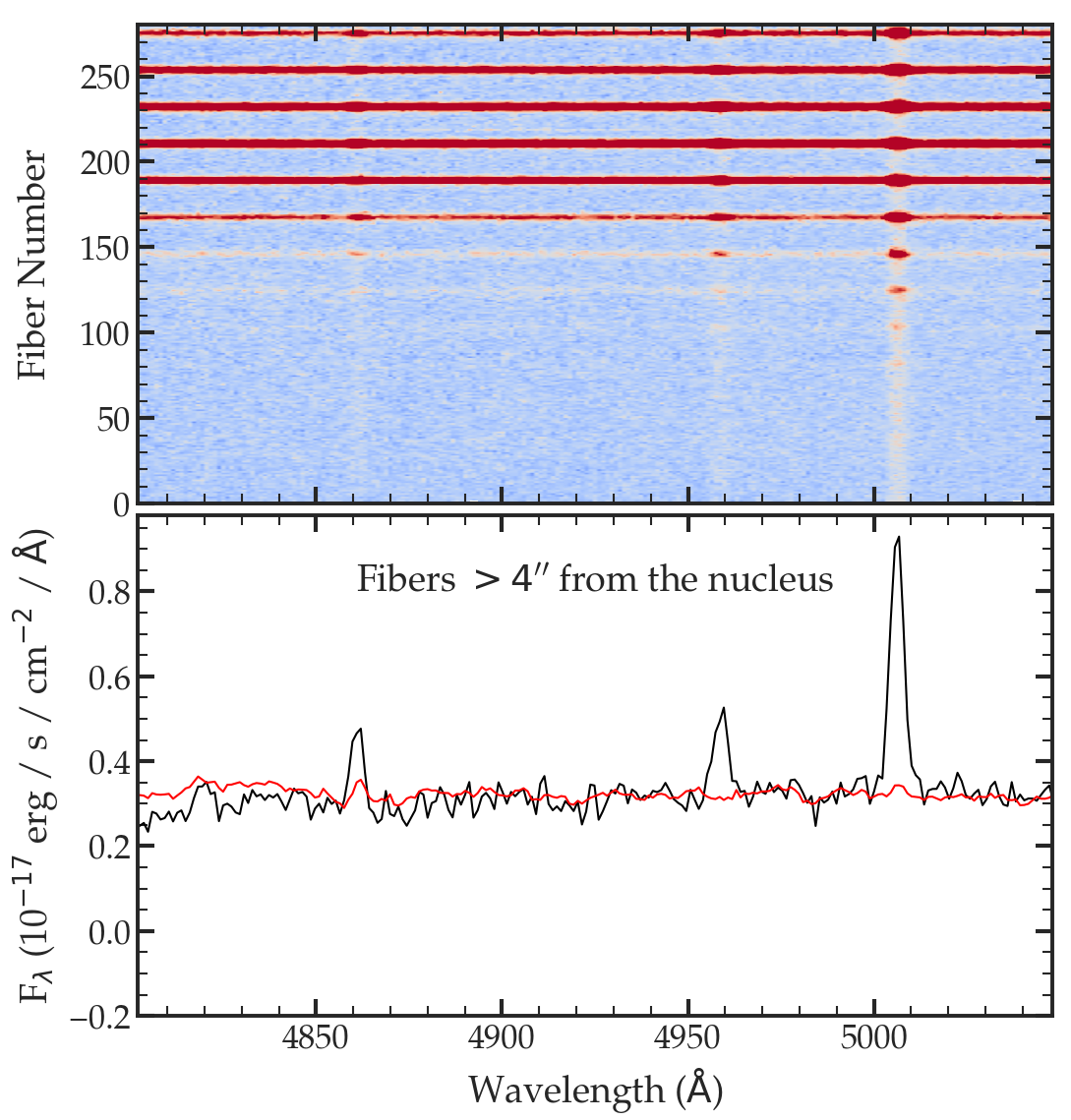}
    \caption{HET LRS2-B detection of the nebula in [\ion{O}{3}]~5007,~4959~\AA\ and ${\rm H\beta}$ emission lines. \textbf{Top Panel:} A demonstrative 2D spectrum, where the vertical axis is the spatial dimension. The emission lines away from the nuclear trace are visible. \textbf{Bottom Panel:} The one-dimensional spectrum constructed in black) from fibers $>4$$''$ away from the nucleus. For comparison, we provide the spectrum of the sky (in red) generated from an observation performed within 30 minutes of the observation of WeSb~1. We note, however, that this reference observation was significantly separated from WeSb~1 spatially, and thus does not capture local spectral features like interstellar nebular emission, if present. Nevertheless, the prominent [\ion{O}{3}] features can be assigned to the WeSb~1 nebula with significant confidence. A prominent ${\rm H\beta}$ emission is also detected, though some fraction of it may arise from the sky/ambient interstellar nebula. }
    \label{fig:nebula_detection}
\end{figure}

The nebula of WeSb~1 is larger than both the slit length of DBSP and the FOV of LRS2-B. Thus, any background subtraction eliminates the nebular features. Faint emission from the surrounding nebula away from the central star is detected in both DBSP and HET images before background subtraction. We present the HET detection (because of its higher signal-to-noise) in Figure \ref{fig:nebula_detection}. We focus primarily on the ${\rm H\beta}$ and [\ion{O}{3}]~4959,~5007~\AA\ emissions, as the sky contamination is minimal in this wavelength range. The extended nature of all three emission lines (due to the nebula) is evident in the two-dimensional spectrum (top panel of the figure). The bottom panel shows a one-dimensional spectrum extracted from fibers sufficiently away from the central core. The emission lines are prominently detected. The line intensity ratio [\ion{O}{3}]~5007~\AA/${\rm H\beta}$ inferred from this spectrum is $\gtrsim$3, consistent with typical values in PNe. Due to insufficient signal-to-noise (or non-detection) of other nebular features and the inability to perform appropriate background subtraction with the current data, we do not perform any quantitative analysis of the nebula in this study. Henceforth, we concentrate on the spectrum of the unresolved central source. 

\subsubsection{Analysis}\label{subsubsec:wesb1_spec_analysis}

\begin{deluxetable}{ccc}
\tablenum{3}
\tablecaption{The inferred fluxes and equivalent widths (EW) for the first three lines of the Balmer-series and the [\ion{O}{3}]~4363,~4959,~5007~\AA\ emission lines for the nucleus of WeSb~1.}
\label{tab:flux_table}
\tabletypesize{\small}
\tablewidth{0pt}
\tablehead{
    \colhead{Line} &
    \colhead{${\rm \log[Flux (erg~s^{-1}~cm^{-2})]}$} &
    \colhead{EW (\AA)}
}
\startdata
${\rm H\gamma}$   & $-14.16\pm0.08$ & $3.13\pm0.58$ \\
 {[}\ion{O}{3}{]}~4363\AA\ & $-13.89\pm0.07$ & $5.58\pm0.92$ \\
 ${\rm H\beta}$   &  $-13.75\pm0.08$ & $5.93\pm1.11$ \\
 {[}\ion{O}{3}{]}~4959\AA\ & $-13.65\pm0.08$ & $7.03\pm1.21$ \\
 {[}\ion{O}{3}{]}~5007\AA\ & $-13.17\pm0.07$ & $21.56\pm2.32$ \\
 ${\rm H\alpha}$ & $-13.03\pm0.02$ & $25.51\pm1.32$ \\
 \enddata
\tablecomments{The reported values are the mean and standard deviation of all the measurements from DBSP and HET spectra. For the Balmer lines, showing both absorption and emission features, we perform two line flux estimations: 1) we fit the profile with two Gaussians and calculate the flux from the emission component, and 2) we use the \texttt{Coelho} model spectrum (inferred in section \ref{subsubsec:wesb1_spec_analysis}) to eliminate the absorption component followed by a single Gaussian fit to the resultant emission profile.}
\end{deluxetable}

\begin{figure}[t]
    \centering
    \includegraphics[width=\linewidth]{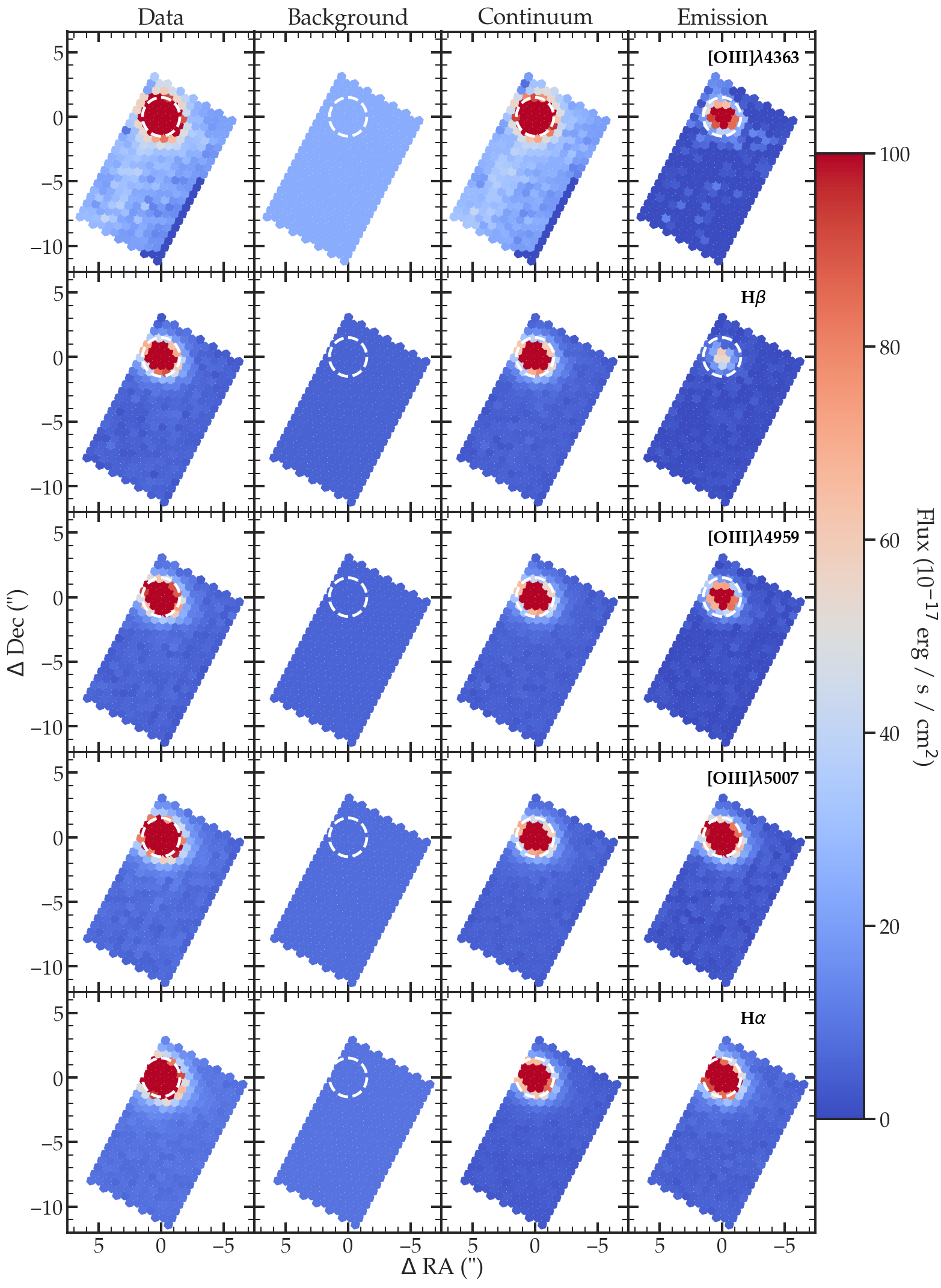}
    \caption{Synthetic narrow-band (NB, with a wavelength full-width of 16~\AA) images of WeSb~1 created from the HET LRS2-B observation on 2024 August 9 around the emission lines of [\ion{O}{3}]~4363,~4959,~5007~\AA, ${\rm H\alpha}$, and ${\rm H\beta}$. The `Data' column shows the NB readings without background\slash nebular subtraction. The second column shows the Background modeled from observations within an annulus of 4-6$''$ from the target. This includes the nebula. The `Continuum' column shows the stellar continuum neighboring each emission line. The final `Emission' column shows the background and continuum subtracted image, showing the compact emission region coincident with the nucleus. The dashed white circles show the 2$''$ radius extraction aperture centered on the nucleus. For further details on the synthesis of NB images from LRS2-B data, refer to sections 4.1 and 4.2 in \cite{Bond24egb6}.}
    \label{fig:HET_2D_image}
\end{figure}

The spectrum of the nucleus shows mostly a flat continuum, with emission lines typical of a planetary nebula: ${\rm H\alpha}$ in the red arm, and ${\rm H\beta}$, ${\rm H\gamma}$, and [\ion{O}{3}]~4363,~4959,~5007~\AA\ in the blue arm. The inferred fluxes and equivalent widths for these emission lines are provided in Table \ref{tab:flux_table}. These emission lines appear compact and confined to the central stellar core. This is evident from both the one-dimensional spatial profile constructed from the DBSP spectra (inset in the bottom panel of Figure \ref{fig:WeSb1_trace_spectrum}) and the two-dimensional narrow-band images from HET (Figure \ref{fig:HET_2D_image}). Most of the observed emission, thus, originates from the central stellar system, which is uncommon. Other prominent emission features identified in the spectrum are [\ion{Ne}{3}]~3869~\AA, \ion{He}{2}~4686~\AA, and \ion{He}{1}~5876~\AA. 

We attempt to estimate the physical properties of the WeSb~1 nucleus using the [\ion{O}{3}] emission line intensity ratios. The following ratio is often used to estimate the electron density and/or temperature: \(\frac{\rm [O~\sc{III}]~5007~\AA+[O~\sc{III}]~4959~\AA}{{\rm [O~\sc{III}]~4363~\AA}}\), which we denote as ${\rm O_{ratio}}$. We use the relation from \cite{Osterbrock04}:
\begin{equation}\label{eq:oratio_equation}
    {\rm O_{ratio}} = \frac{7.9e^{3.29\times10^4/T_e}}{\left[1+(4.5\times10^{-4}n_e/\sqrt{T_e})\right]},
\end{equation}
which holds reasonably in most cases. The typical values of ${\rm O_{ratio}}$ for PN are $\sim$50 -- 100 which results from electron temperatures of order ${T_{\rm e} = 10^4}$~K and densities of ${n_{\rm e} = 10^{3-4}~{\rm cm^{-3}}}$. The inferred ${\rm O_{ratio}}$ value for WeSb~1 is, however, $\sim$$7$, which is significantly low (with extinction correction expected to make the ratio even smaller) and indicates very high electron densities of order $10^6~{\rm cm^{-3}}$ or higher. For example, with physically realistic values of ${T_{\rm e}\sim1-2\times10^4}$~K, we get ${n_{\rm e}\sim1-5\times10^6~{\rm cm^{-3}}}$. This fact is also supported by the absence of emission lines like [\ion{O}{2}] and [\ion{N}{2}] which likely have attained saturation below the detection limit owing to their low critical densities. At such high densities, the [\ion{O}{3}]~4959,~5007~\AA\ emission lines also near saturation (making Equation \ref{eq:oratio_equation} not as good an approximation anymore) and may result in under- and over-estimation of  ${\rm O_{ratio}}$ and ${n_{\rm e}}$ respectively. A more thorough analysis of the different spectral features is underway; however, it is unlikely to modify the inference significantly.\footnote{We contrast this with the nebular properties. Though not shown in Figure \ref{fig:nebula_detection}, we did not detect [\ion{O}{3}]~4363~\AA\ in the current nebular spectrum. Thus, the nebular ${\rm O_{ratio}}$ is likely very high, implying a low $n_e$, consistent with an evolved nebula like WeSb~1. This also shows that the detected nebular emission lines are ``real" and not an effect of large seeing.}

We proceed to estimate the total line of sight extinction to the central star using the Balmer emission line ratios (Balmer decrement, henceforth BD). We first use the ratio of ${\rm H\alpha}$ to ${\rm H\beta}$, which we denote as ${\rm BD_{\alpha\beta}}$. Assuming a Case B recombination scenario, a temperature of ${1-2\times10^4}$~K, and electron density of ${\rm 10^6~cm^{-3}}$ (as inferred previously), we expect a theoretical value of ${\rm BD_{\alpha\beta,th}} \sim 2.8$ \citep{Osterbrock04}. The inferred ratio from the observed emission fluxes is ${\rm BD_{\alpha\beta, obs}} = 5.25\pm0.76$. The ratio of the observed and theoretical values of ${\rm BD_{\alpha\beta}}$ is related to the net reddening as: ${E_{B-V}} = {\rm 1.97\log\left({\rm BD_{\alpha\beta, obs}}/{\rm BD_{\alpha\beta, th}}\right)}$ (see for example \citealt{Dominguez13}). Using this relationship and ${R_{V} = 3.1}$, we get an estimate of ${A_{V} = 1.64\pm0.4}$.

We perform a second estimation, using ${\rm H\beta}$ and ${\rm H\gamma}$. We broadly follow the method employed in \citet{Aller15}. At the same physical conditions as previously, the theoretical line flux ratio ${\rm BD_{\beta\gamma,th} = H\beta/H\gamma} = 2.11$. The observed flux ratio is ${\rm BD_{\beta\gamma,obs} = 2.57\pm0.4}$. Using the prescription laid out in \citet{Seaton79}, we obtain the relation ${E_{B-V}} = {\rm 5.36\log(BD_{\beta\gamma,obs}/BD_{\beta\gamma,th})}$. This yields ${ A_{V} = 1.34\pm1.14}$, which is inconclusive, but still consistent with the previous estimate. 

The spectrum also displays several absorption features, likely of stellar origin. Hydrogen absorption is detected as broad features at the base of the emission lines. Absorption is more prominent in ${\rm H\beta}$ and the higher-order Balmer series on the blue arm, and the Paschen series on the red. Prominent metal absorption lines are also detected. A strong \ion{Ca}{2}~K~3933~\AA\ absorption line is evident, with a large equivalent width (EW) of $5.5\pm0.5$. The presence of a \ion{Ca}{2}~H~3969~\AA\ cannot be confirmed due to blending with the adjacent Balmer features. \ion{Ca}{2}~triplet~8498,~8542,~8662 \AA\ absorption are also seen but blended with the adjacent Paschen lines. A \ion{Na}{1}~D doublet absorption (at 5890 and 5896 \AA, unresolved in the spectrum) is also present. To obtain the EW, we fit the adjacent \ion{He}{1}~5876~\AA\ emission and the \ion{Na}{1}~D absorption simultaneously with two Gaussian profiles. We obtain a \ion{Na}{1}~D1+D2 combined EW of $1.8\pm0.1$~\AA. Assuming this to be solely extra-stellar origin, and using the relationship between the EW and ${E_{B-V}}$ as given in the bottom panel of Figure 9 in \citealt{Poznanski12}, we get $A_V$$>$$4$. This is significantly higher than the previous estimates. However, we suspect a significant stellar contribution to this absorption which is also supported by the analyses presented next.

\ion{Ca}{2}~triplet (Paschen series) absorption is usually absent in stars with high (low) temperatures. The presence of both features in this spectrum alone constrains the stellar temperature between $\sim$5000~K and $\sim$9000~K. We queried the high-resolution theoretical stellar spectra from the library of \citet[][hereafter, \texttt{Coelho} spectra]{Coelho14} in the above temperature range, at an interval of 500~K from the SVO theoretical services website\footnote{\url{http://svo2.cab.inta-csic.es/theory/main/}}. For simplicity, we restricted to parameters ${\log(g/{\rm [cm~s^{-2}]})=3}$ (this choice is also influenced by the inferred large radius of the star, as described later) and [Fe/H] = 0. With $R$$\sim$1500, the spectral resolution of DBSP is $\Delta\lambda$$\sim$$4$~\AA. We downgraded the \texttt{Coelho} spectra to DBSP resolution by convolving with a Gaussian kernel with an FWHM of 200 pixels (4~\AA) and performed a visual comparison of the spectral features.

\begin{figure*}[t]
    \centering
    \includegraphics[width=\linewidth]{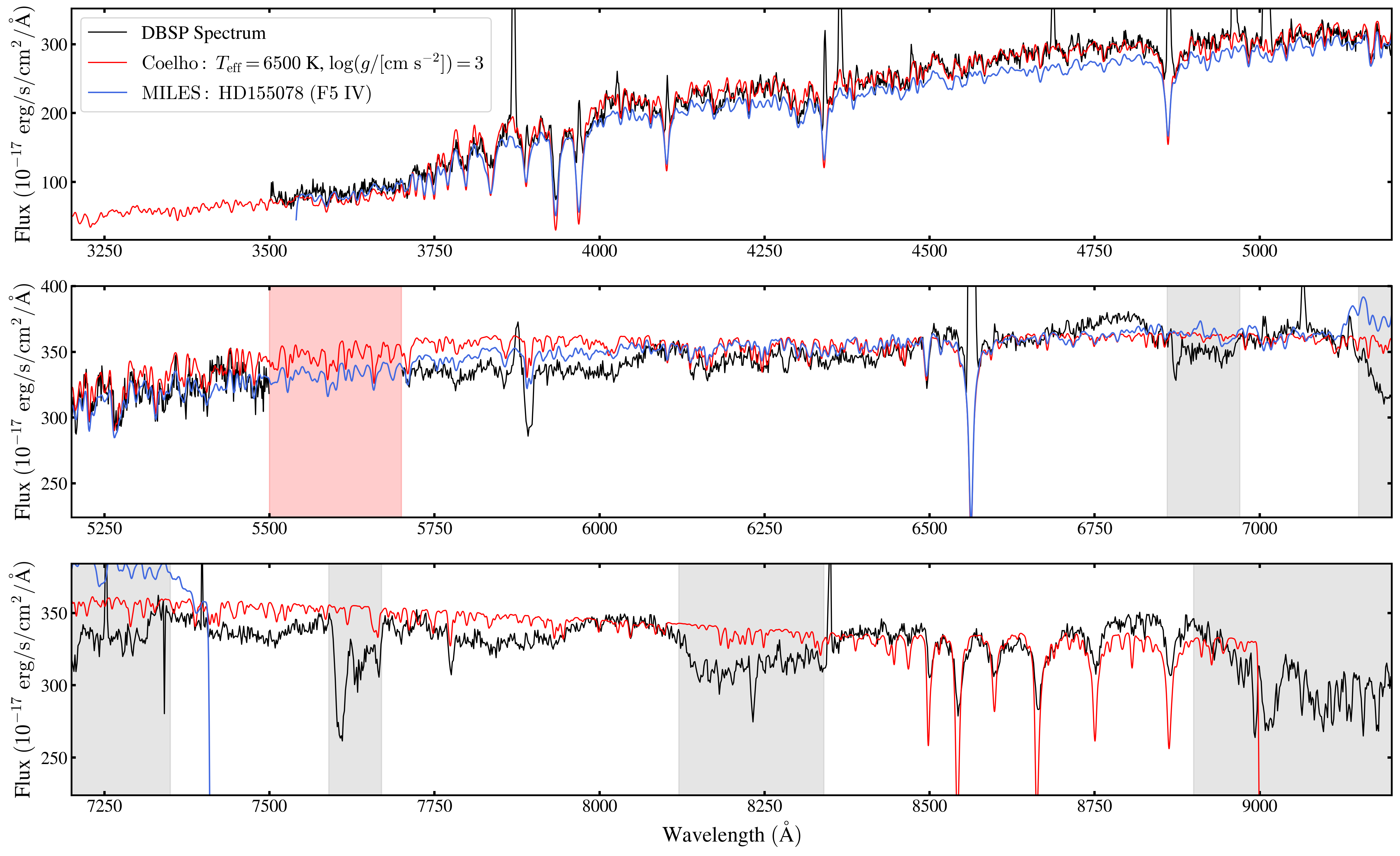}
    \caption{Comparison of the DBSP spectrum (black) with synthetic \texttt{Coelho} spectrum (red) and that of the F5~type subgiant HD\,155078 (from MILES survey, blue). The entire wavelength range is subdivided into three parts, and presented in three panels, each spanning 2000~\AA. The MILES spectrum, unfortunately, is not available from $\sim$7400~\AA\ onwards. The \texttt{Coelho} spectrum truncates only at 9000~\AA, thus covering all the essential spectral features. We omit the portion of the DBSP spectrum (shaded in red) that marks the transition from the blue to the red arm (and is thus extremely noisy). We also shade in gray the regions of strong telluric features (by referring to \citealt{Rudolf16}).}
    \label{fig:WeSb1_spec_compare}
\end{figure*}

A satisfactory match was obtained for ${T_{\rm eff}=6500}$~K (for which we assign an uncertainty of $\sim$500~K, which is the interval for comparison, can be assumed). The result is presented in Figure~\ref{fig:WeSb1_spec_compare}. The best match to the overall spectral shape was achieved with a reddening of ${A_V\simeq1.9}$\footnote{All dust extinction corrections in this section have been performed using the \texttt{dust\_extinction} \citep{dust_extinction, Gordon24_package_paper} implementation of the \cite{Karl23} galactic extinction models \citep{Decleir22, Gordon21, Fritzpatrick19, Gordon09}.}, in reasonable agreement with that inferred from the Balmer lines (especially with ${\rm BD_{\alpha\beta}}$). The synthetic spectrum has been scaled to match the observed flux. The value of the scaling factor corresponds to a stellar radius of ${R_{*}}\simeq6\,{R_{\odot}}$\footnote{The scaling factor ${=\pi(R/D)^2}$, where ${R}$ and ${D}$ are the stellar radius and distance, respectively. The factor of $\pi$ comes from the angular integration over a Lambertian surface.} (when the \emph{Gaia} distance of 3.7~kpc is assumed). At the inferred temperature, this radius is more consistent with a subgiant star. For further comparison, we queried the MILES archival spectral database from the SVO portal for a star of suitable parameters. The spectral resolution of MILES observations is $\sim$$1.5$ times that of DBSP, thus we down-resolve the data by convolving with a Gaussian kernel of width 2 pixels (1.8\AA). The F5~IV-type star,\footnote{This is a visual demonstrative comparison and may not be regarded as a spectral type identification} HD\,155078, with ${T_{\rm eff}}=6418$~K, $\log(g/{\rm [cm~s^{-2}]})=3.71$, and $\rm[Fe/H]=-0.1$ yielded a good match with both the \texttt{Coelho} spectrum and DBSP (shown in the same figure). The absolute \emph{Gaia} $G$ magnitude for this reference star is $M_G=+2.2$, close to that of WeSb~1 ($M_G = +1.98$). A star with this low a  temperature cannot produce high-excitation emission lines like [\ion{O}{3}] and \ion{He}{2} as seen in the spectrum. This strongly suggests the presence of a hot CSPN, with the cool star being its companion.

Though the overall agreement between the model and DBSP spectra is good, there are differences in some of the detailed features. The \ion{Ca}{2}~K absorption in the blue and the \ion{Ca}{2}~triplet and Paschen series absorption in the red, for example, appear stronger in the models. We emphasize here that the spectral comparison was performed only visually and, thus, a perfect match is not expected. A more quantitative fit will be performed in the future paper on this object. The match between the models and the DBSP spectrum is also better in the shorter wavelengths (the top panel in Figure \ref{fig:WeSb1_spec_compare}) than the rest of the spectrum. This is possibly due to the contamination from circumstellar dust emission (see next section) and the plethora of atmospheric features in the longer wavelengths.

\subsection{Spectral Energy Distribution}\label{subsubsec:wesb1_sed}

\begin{deluxetable}{ccc}
\centering
\tablenum{ 4}
\tablecaption{The flux outputs from \texttt{VOSA} used to construct the SEDs in Figure \ref{fig:wesb1_sed}. The photometric bands of the various surveys are listed in the first column. The second column provides the respective central wavelengths (in angstroms). The last column provides the observed flux values in units of ${\rm 10^{-15}~erg~cm^{-2}~s^{-1}~\AA^{-1}}$}.
\label{tab:wesb1sed}
\tabletypesize{\footnotesize}
\tablewidth{0pt}
\tablehead{
    \colhead{Photometric Bands} &
    \colhead{Wavelength} & 
    \colhead{Obs. Flux} 
}
\startdata
GALEX/GALEX.NUV    & $2303.37$                & $0.225\pm0.021$  \\
Misc/APASS.B       & $4299.24$                & $2.806\pm0.370$    \\
Generic/Johnson.B  & $4369.53$                & $3.120\pm0.023$     \\
SLOAN/SDSS.g       & $4671.78$                & $3.567\pm0.234$    \\
PAN-STARRS/PS1.g   & $4810.16$                & $2.838\pm0.127$     \\
GAIA/GAIA3.Gbp     & $5035.75$                & $3.210\pm0.060$    \\
Misc/APASS.V       & $5393.85$                & $3.617\pm0.356$    \\
Generic/Johnson.V  & $5467.57$                & $3.833\pm0.019$    \\
HST/ACS\_WFC.F606W & $5809.26$                & $3.725\pm0.014$     \\
GAIA/GAIA3.G       & $5822.39$                & $3.090\pm0.019$     \\
SLOAN/SDSS.r       & $6141.12$                & $3.859\pm0.159$     \\
PAN-STARRS/PS1.r   & $6155.47$                & $3.180\pm0.075$      \\
Generic/Johnson.R  & $6695.83$                & $3.583\pm0.016$      \\
SLOAN/SDSS.i       & $7457.89$                & $3.109\pm0.133$      \\
PAN-STARRS/PS1.i   & $7503.03$                & $2.622\pm0.082$      \\
GAIA/GAIA3.Grp     & $7619.96$                & $2.934\pm0.043$      \\
HST/ACS\_WFC.F814W & $7973.39$                & $2.912\pm0.014$      \\
Generic/Johnson.I  & $8568.89$                & $2.925\pm0.015$      \\
PAN-STARRS/PS1.z   & $8668.36$                & $2.772\pm0.039$      \\
SLOAN/SDSS.z       & $8922.78$                & $2.714\pm0.018$     \\
PAN-STARRS/PS1.y   & $9613.60$                & $2.533\pm0.170$      \\
2MASS/2MASS.J      & $12350.00$               & $2.221\pm0.049$      \\
2MASS/2MASS.H      & $16620.00$               & $1.822\pm0.025$      \\
2MASS/2MASS.Ks     & $21590.00$               & $1.474\pm0.029$      \\
WISE/WISE.W1       & $33526.00$               & $0.908\pm0.018$      \\
WISE/WISE.W2       & $46028.00$               & $0.572\pm0.010$      \\
WISE/WISE.W3       & $115608.00$              & $0.098\pm0.001$      \\
WISE/WISE.W4       & $220883.00$              & $0.031\pm0.001$   \\
\enddata

\end{deluxetable}

\begin{figure}[t]
    \centering
    \includegraphics[width=\linewidth]{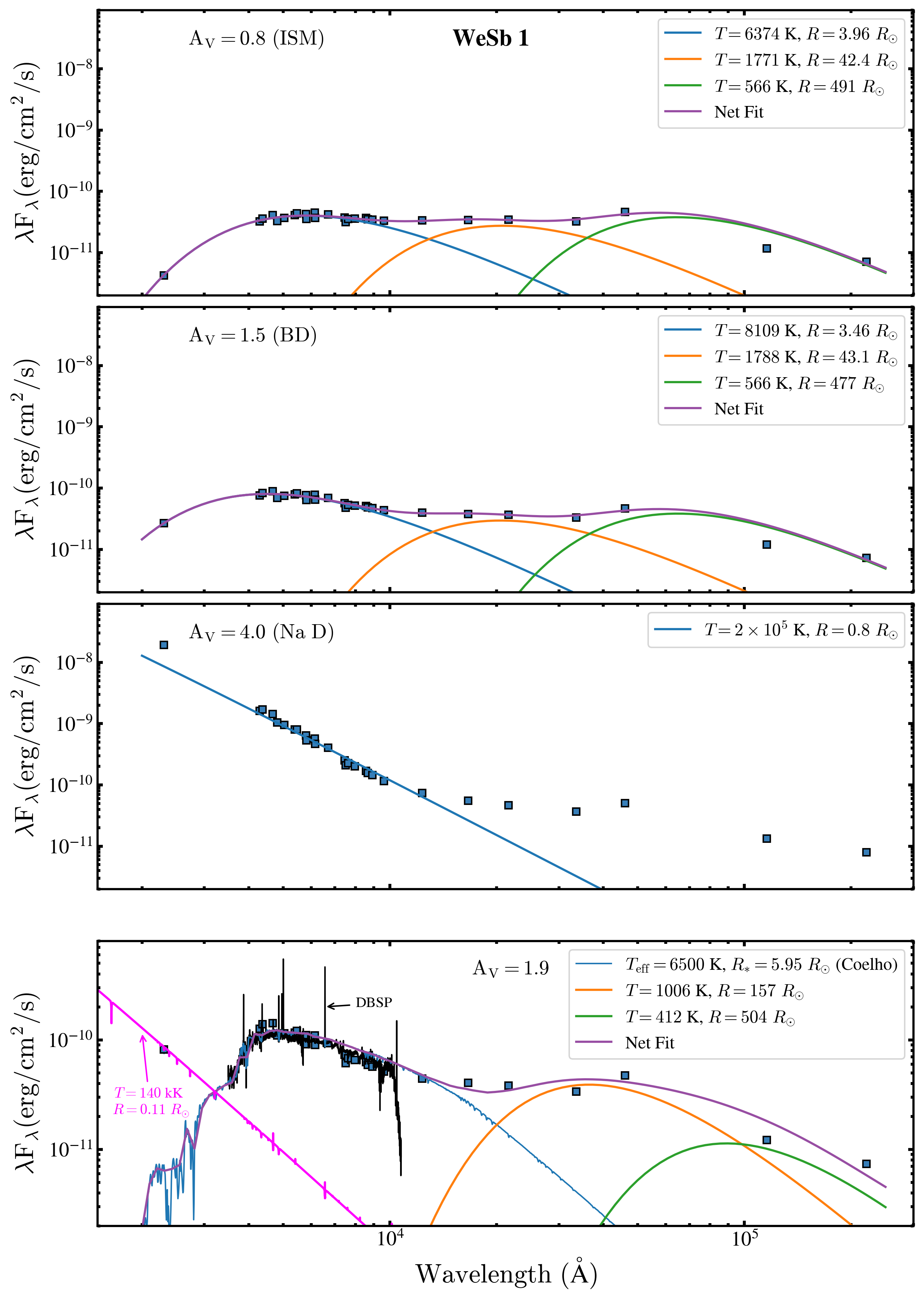}
    \caption{\textbf{Top three panels:} The de-reddened SEDs for WeSb~1, with the three cases of extinction (noted within the figure). The empirical three-blackbody fit (both the individual components and the net fit, see the color code in the legend) to the SEDs are also shown, except for the case of ${A_V = 4.0}$, where a manually chosen exemplary fit to the optical data points is shown. \textbf{Bottom panel:} De-reddened SED with ${A_V=1.9}$ with the DBSP and \texttt{Coelho} spectra overplotted. The two-blackbody fit to the longer wavelengths are shown. An example hot blackbody component that fits the GALEX NUV data points is also shown.}
    \label{fig:wesb1_sed}
\end{figure}

We queried the SED of WeSb~1 using the online \texttt{VOSA} SED tool \citep[][Figure \ref{fig:wesb1_sed}, blue squares]{Bayo08} with a 3$''$ cross-match radius with all the available catalogs. The flux outputs are presented in Table \ref{tab:wesb1sed}. \texttt{VOSA} failed to load the GALEX near-ultra-violet (NUV) flux, which was separately queried from the GALEX catalog\footnote{\url{https://galex.stsci.edu/GR6/?page=mastform}} and manually added to the table. 

First, we analyze the SED independent of the inferences about the stellar parameters/spectral type in the previous section. We consider three cases of extinction: only interstellar (${A_V=0.8}$\footnote{Obtained from the online query service: \url{http://argonaut.skymaps.info/query}, implementing the \cite{Green19} three-dimensional extinction maps}), a representative value for the Balmer decrement (${A_V=1.5}$), and the Na-D absorption (${A_V=4.0}$, as an example of an extreme case of extinction).\footnote{We note here that the EW of the Na-D absorption in the \texttt{Coelho} spectrum is $\sim$$0.5$. Subtracting this from the observed EW leads to a revised extinction estimation of $\sim$$1.5\pm0.4$, consistent with all the other estimates. We still briefly consider this case just to demonstrate the effect of a large extinction, in case it is underestimated.} The first three panels of Figure \ref{fig:wesb1_sed} show the reddening-corrected SEDs for these three extinction values. A strong IR excess, starting from the near-infrared wavelengths is evident, likely arising from circumstellar dust disk and\slash or shell. The stellar component dominates in the optical wavelengths.

To extract quantitative information, we attempt to fit the SEDs with blackbody functions. A reasonable fit was achieved with three blackbodies except for the case of ${A_V=4.0}$, where the fit did not converge. The data points for wavelengths $\gtrsim$$1~{\rm \mu m}$ (dust components) are well fit by two blackbody components. The first component is relatively hot and compact with $T$$\sim$$1800$~K and a radius of $\sim$$40~{R_{\odot}}$. The second component is more diffused with $\sim$$200-400~{\rm R_{\odot}}$ and $T$$\sim$$600$~K. 

The data at the shorter wavelengths (stellar component) are well-fit by a single blackbody. The inferred temperature, however, is dependent on the extinction value assumed. Overall, the inferred temperature and radius are in the range of $6000-8500$~K and $3-4~{R_{\odot}}$, respectively. These parameters are broadly consistent with stars between spectral types A and G (consistent with the inferences in the previous section). We do not quote a value for the case of ${A_V=4.0}$, but a very high temperature is required to fit the data. Shown in the Figure is a manually chosen example of ${T\sim2\times10^5~K}$ and ${R\sim0.8R_{\odot}}$. Very hot [WR] stars come closest to these parameters, which may be the CSPN itself. However, this extinction scenario, as mentioned earlier, is unlikely.

We now use the inferences from the previous section and analyze the SED. The results are presented in the last panel in Figure \ref{fig:wesb1_sed}. We use ${A_V=1.9}$ (which yields the best visual match between the synthetic and observed spectrum, see previous discussion) to de-redden the SED. We overplot the DBSP and the synthetic \texttt{Coelho} spectra. For the latter, we queried the low-resolution version, owing to its much longer wavelength coverage. The strong IR excess is still evident. To perform the model-fit, we assume the \texttt{Coelho} spectrum for the stellar component, keeping the radius as a free parameter, and model the longer wavelength data points as two black bodies. The resultant fits are shown in the same figure. The inferred radius of the stellar component is ${5.95~R_{\odot}}$ (the uncertainty in the \emph{Gaia} distance translates to a $\sim$$0.5~R_{\odot}$ uncertainty here). Here, too, two dust components are seen. The inferred temperature of the warmer component, however, is lower than those inferred in the previous analyses with three black bodies.

The notable feature in this case is the appearance of a prominent NUV excess, possibly from the hot CSPN. With only a single data point, it is not possible to get a good constraint on the properties. But, for any reasonable assumption of the temperature, a radius $>$$0.05~R_{\odot}$ is required to match the observed NUV flux. This can be either a (still contracting) very young white dwarf or a [WR] star. As an example `fit', we over-plot the non-local-thermodynamic-equilibrium (NLTE) atmosphere model of a hot star (with a solar composition of H and He). Further observations (especially an ultraviolet spectrum) are needed to infer the nature of the hot CSPN.

\subsection{Discussion}\label{subsec:wesb1_discussion}

As mentioned earlier, the irregular dipping behavior resembles systems showing transits from dust or debris discs. Being the central star of a likely PN, we relate the variability to white dwarfs with transiting planetary debris discs. In the late to post-AGB phase, stars can perturb the orbits of their planetary systems to highly eccentric orbits. This leads to tidal disruption of the rocky bodies, leading to a debris disk around the young white dwarf. Recently, several white dwarfs showing transiting planetary debris have been discovered \citep{Vanderburg15, Vanderbosch20, Guidry21}, with very similar light curve behavior. WeSb~1 stands as a candidate for such a system in formation. 

The position of the star in the \emph{Gaia} color-magnitude diagram (CMD), however, is not consistent with a single white dwarf. With \emph{Gaia} $G_{\rm BP}-G_{\rm RP}=1.17$ and ${G_{\rm abs} \sim 2}$, the object is significantly redder and more luminous than a white dwarf. All the previous analyses substantiate this, making it more likely to be a binary with the optical flux dominated by a cooler companion (A -- G type star) of the hot CSPN (a chance alignment of a cool field star and the hot CSPN is unlikely as the probability of such an occurence is very low, see Appendix \ref{appendix:misidentified}). Thus, the observed photometric variability can arise from a transiting circumbinary debris or a circumstellar debris disk around the companion. The near-total eclipses, however, demands that the obscuring material be nearly as large as the star itself. With the inferred stellar radius being $\sim$$6~R_{\odot}$, it is more likely that the transits are from very thick dust/debris clouds rather than solid objects like asteroids. We note here that dust activity from the star itself might also have resulted in irregular variability (with RCrB stars being extreme cases\footnote{Note that the WeSb~1 nucleus is unlikely to be an RCrB or similar hydrogen deficient star due to the presence of prominent Balmer\slash Paschen absorption lines}). However, the best estimate of the spectral type of the companion, F-type subgiant, makes this scenario unlikely as such stars are not expected to take part in significant dust production. 

\begin{figure*}[t]
    \centering
    \includegraphics[width=\linewidth]{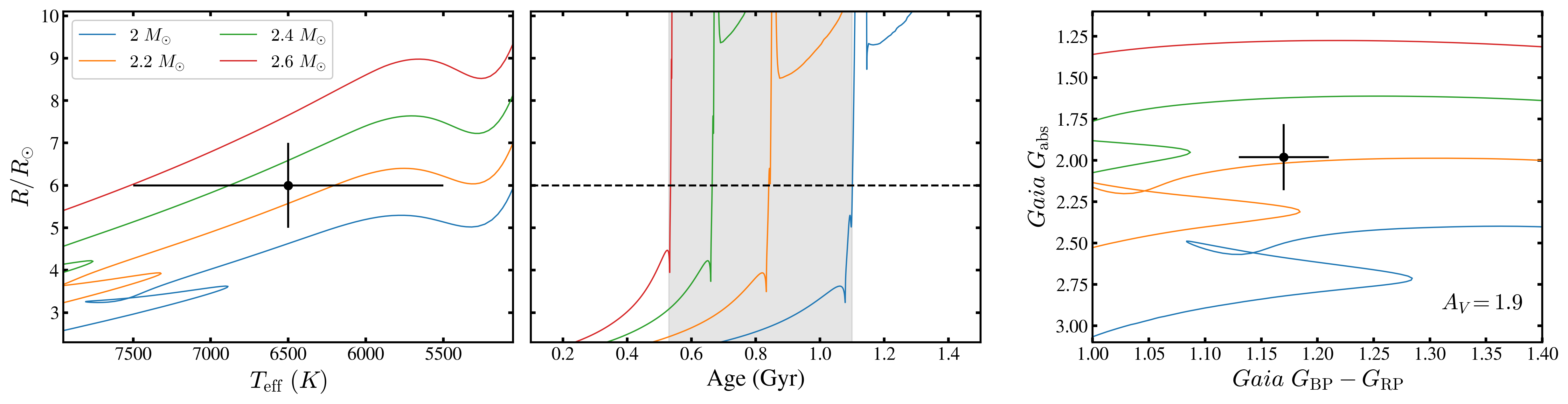}
    \caption{Comparison of the observed/inferred stellar parameters (black points) to MESA evolutionary models. Four stellar masses yielding the closest match to WeSb~1 are shown (colored lines). \textbf{Left panel:} The evolutionary tracks in the space of radius and temperature. For the inferred properties we use ${T_{\rm eff}=6500}$~K and ${R_*=6~R_{\odot}}$ (see \ref{subsubsec:wesb1_spec_analysis}). For a generous estimate, we have doubled the uncertainties in these parameters (from those discussed in sections \ref{subsubsec:wesb1_spec_analysis} and \ref{subsubsec:wesb1_sed}) to $1000$~K and $1~R_{\odot}$, respectively. \textbf{Middle Panel:} The evolution of stellar radius with age. We only plot the regime where $T_{\rm eff}<15{,}000$~K to avoid confusion with the post-AGB track. The horizontal line marks the median inferred radius, and the shaded region depicts the age range estimated from the four stellar masses. \textbf{Right Panel:} \emph{Gaia} CMD. Extinction of ${A_V=1.9}$ was applied in querying the model tracks.}
    \label{fig:wesb1_mist_compare}
\end{figure*}

The inferred temperature and radius of the companion can be used to estimate its mass and the age of the system. We query stellar evolutionary tracks calculated using the Modules for Experiments in Stellar Astrophysics (MESA, \citealt{Paxton11,Paxton13,Paxton15}) from the online portal of MESA Isochrones and Stellar Tracks (MIST, \citealt{Dotter16, Choi16})\footnote{\url{https://waps.cfa.harvard.edu/MIST/index.html}}. We visually compare the models with the stellar parameters calculated in the previous sections. The result is presented in Figure \ref{fig:wesb1_mist_compare}. We find that the inferred radius and temperature values lie between stars of masses 2.2 -- 2.4~$M_{\odot}$ (left panel in the figure). With a generous allowance for error, we arrive at a mass estimation of ${2.3\pm0.3~M_{\odot}}$. Thus, we get a corresponding age of the star of ${\rm 0.75\pm0.25}$~Gyr (middle panel in the figure). Note that at this evolutionary stage, the stellar radius (and temperature) evolves rapidly with time. Thus, the age estimation is not particularly sensitive to the exact estimations of the stellar parameters. At this mass and age, the model predicts ${\log(g/{\rm [cm~s^{-2}]})\sim3.2}$, consistent with the previous estimates. The position of the star in the \emph{Gaia} CMD is also consistent with those predicted by the models in the inferred mass range (with an applied extinction correction of ${A_V=1.9}$, see right panel of the same figure). As the PN is still visible, this implies that the PN progenitor left the main sequence only recently. Thus, its initial mass needs to be very similar (slightly larger) to its subgiant companion. We thus obtain estimates of both the total initial mass of the binary and its age. We also find that the example stellar parameters of the hot CSPN used in Figure \ref{fig:wesb1_sed} are also broadly consistent with this evolutionary stage (see Appendix \ref{appendix:wesb1_cspn_mist_compare}). We note here, however, that there is a chance that the companion is not naturally in the subgiant phase (as indicated by the radius), but is just inflated from capture of the nebular ejecta. In that case, the mass and age estimation will no longer be valid.

The compactness of the emission lines and the high inferred electron densities (see section \ref{wesb1_spectrum}) are additional puzzles. These suggest a high-density emitting core around the WeSb~1 nucleus, making it a double-envelope nebula. Such a configuration bears a close resemblance to a rare class of PN, where the central core hosts a high-density compact emission knot (CEK). The formation channels of CEKs are not yet well understood. The leading hypothesis is the fall-back of some of the nebular ejecta on the white dwarf or its wide-orbit companion. The prototypical system is EGB~6 \citep{Liebert13, Bond16}, with only around a dozen more known of its kind (see \citealt{Frew16}, and the recent discoveries by \citealt{Bond24egb6, Bond24egb62}). The high inferred electron densities, near-IR excess, and the presence of a companion in WeSb~1 are all consistent with the properties of a EGB~6-type PN nucleus.

We now consider a different possibility of WeSb~1 being a symbiotic system, where such high electron densities and compact emissions are common. The BPT-like diagnostic diagram based on the [\ion{O}{3}] and Balmer emission line ratios (often used to distinguish PN from the symbiotic systems, see for example \citealt{Gutierrez-Moreno95}) places the source in the regime of D-type (Mira donor) or D'-type (G or K-type giant donor) symbiotic star (though, note that such an apparent position can also result with CEKs). In most D-type symbiotic systems, the observed nebula is formed from the ionized winds of the Mira giant, and thus not a true PN (see for example discussion in \citealt{Frew16}). WeSb~1 is unlikely to be a D-type symbiotic due to i) the lack of the near IR peak in the SED from the warm dust expected in such systems, and ii) the absence of Mira signatures in the spectrum. In any other symbiotic system, the donor winds are not expected to be strong enough to create the nebula, thus any observed nebula is a true PN from the accretor progenitor. Only three such systems are currently known. If WeSb~1 happens to be such a system, we conjecture that the subgiant companion is filling its Roche-lobe and transferring mass to the hot white dwarf. The formal symbiotic category closest to this scenario is D'-type symbiotic. Flat SED from optical through near-IR, a significant mid-IR excess, and the requirement to use three black body components are indeed characteristics of D'-type symbiotic systems (see discussion in section 3 and figure 4 in \citealt{Akras19}). In such a scenario, for any reasonable mass of the accretor ($\simeq$$0.7~{M_{\odot}}$) and with the estimated mass of the companion, the period then should be $\lesssim$$3$~days. A future radial velocity analysis will test this scenario. Concerning the current data, no confirmatory symbiotic spectral signatures like Raman-scattered \ion{O}{6} emission lines at wavelengths 6825, 7082~\AA, are detected in the current spectrum. The object's position in diagnostic diagrams based on 2MASS and WISE colors (see for example Figure 1 and Figure 2 in \citealt{Stavros19}) are inconclusive in this regard.

We compare WeSb~1 with other systems known to show dust obscuration and find a similarity to NGC~2346. It is a binary system with one of the longest known periods inside a PN (16~days, see \citealt{Mendez1981, Brown19}) and known to show dust obscuration \citep{Mendez82}. The binary consists of a hot CSPN and an A-type subgiant (see \citealt{Gomez19}) with stellar parameters comparable to those estimated for WeSb~1 in this work. Unlike WeSb~1, however, NGC~2346 does not show a high-density emission-line core, nor do they have the same nebular morphology (the latter is bipolar). We note here though that it is possible for WeSb~1 nebula to be bipolar, with the viewing angle being face-on. However, this poses an additional challenge of explaining a polar dust disk which can cause the transits. WeSb~1 also bears similarities to PN~M~2-29 \citep{Hajduk08, Miszalski11} which show both dust obscuration and an emission-line core. In fact, the latter work discusses the same two possibilities for PN~M~2-29 (EGB~6-type PN or D'-type symbiotic) as has been discussed in this work for WeSb~1. However, one difference remains: the transits in WeSb~1 are much shorter-lived (a few days) compared to the others (a few years).

\section{Discussion and Conclusions}\label{sec:conclusion}

We have conducted a systematic study of the optical photometric variability of cataloged central stars of planetary nebulae (CSPNe, \citetalias{Chornay21Distance}) using the light curves from the Zwicky Transient Facility (ZTF). The main challenge in studying PNe with ground-based telescopes is atmospheric seeing. This often prevents adequate resolution of the CSPN from the surrounding nebula, introducing uncertainties (see Section \ref{subsec:photometry}). We employed stringent photometric quality cuts (Section \ref{subsec:quality_cuts}) and used appropriate variability metrics (Normalized Excess Variance, see Section \ref{subsec:metrics_and_period}) to mitigate such effects. We arrived at a final list of $94$ highly variable CSPN candidates (the HNEV sample). The variables can broadly be classified into two qualitative classes: short-timescale and long-timescale, based on the temporal resolution of the variability with ZTF cadence. In this paper, we focus on the former class. The latter is the subject of Paper II (in prep.). The list of short-timescale variables is presented in Table \ref{tab:short_timescale_vars}.

Most (83) of the HNEV CSPNe belong to the class of short-timescale variables. We searched for periodicity using the Lomb-Scargle algorithm (Section \ref{subsec:short_timescale_periodic}). We recovered seventeen known periodic variables. Among them, the correct period was detected in twelve and half the true period in another three. We also find six new periodic sources, with confident period detection in five of them (also see Appendix \ref{appendix:future_prospects} for three more sources among the non-HNEV objects). Among the aperiodic light curves, the majority are jittering sources showing irregular variability with flux amplitude in the range of $\sim$10 -- 40\%. Possible sources of such variability include wind from the hot CSPN, unresolved pulsations and/or local stochastic dust activity. We also find several outbursting sources. A literature search shows that we have recovered several objects with prior reports of photometric variability, including a significant fraction of the \emph{Gaia} variables reported in \cite{Chornay21Binary}.

The variability in one of the objects, WeSb~1, appears unique. The optical light curves show very deep and irregular dips that resemble transits from debris disk (see Figure \ref{fig:wesb1_ztf_gattini_ir}). The ingress and egress durations of the transits range over a few days with short-duration minima. The transits are well-resolved in the available TESS light curves. Similar behavior is also witnessed in the near-IR $J$-band light curve from Palomar Gattini-IR. Significant variability is also seen in \emph{Gaia} and WISE light curves, though their cadence is inadequate in resolving the short-timescale activities. Such peculiarity prompted additional studies (see section \ref{appendix:wesb1}). We find strong evidence for the binarity of the system. The optical spectrum and the SED constrain the spectral type of the companion roughly between A and G, with the best estimate being an F-type subgiant. With this knowledge, we use stellar evolutionary models to constrain the initial mass and the age of the system (see Section \ref{subsec:wesb1_discussion} and also Appendix \ref{appendix:wesb1_cspn_mist_compare}). The SED shows significant IR excess starting from near-IR wavelengths, indicative of a complex dust structure around the nucleus. Most of the line emissions appear confined to the nucleus, with the line ratios suggesting very high electron densities ($\gtrsim$$10^6~{\rm cm^{-3}}$). This is unexpected given the evolved morphology of the surrounding nebula. This makes WeSb~1 a candidate for either EGB~6-type PN hosting compact emission knots, or a symbiotic system inside an evolved PN, both of which are rare systems. A more detailed study on this object with additional data is in preparation.

We conclude by briefly discussing the prospects of using ground-based photometric sky surveys to study CSPNe. With much higher cadence and longer temporal baselines compared to most sky-telescopes, these are ideal for studying the photometric variability of sources and resolving small timescale activities. This work (Paper I) and Paper II (in prep.), demonstrate the success of the existing facilities (primarily ZTF, but also other surveys like Asteroid Terrestrial-impact Last Alert System (ATLAS) in optical and Palomr Gattini-IR in near-IR) to identify unique sources. The upcoming Large Synoptic Survey Telescope (LSST, \citealt{Ivezic19}) will widen the window of opportunity with its access to the galactic center where most of the PNe are situated. Effective variability metric(s) need to be identified to find `interesting' sources from the pool of data (see Appendix \ref{sec:vonn_skewp} for such an example of a two-dimensional metric space). However, most algorithms currently in existence focus on point-source photometry. With the extended (and often bright) nebula, this is unsuitable for the study of CSPNe. On the other hand, simple aperture photometry of the whole nebula is also not desirable as i) we are interested in the CSPN, and ii) contamination from several field stars in the galactic plane. Thus, sufficient care must be taken in using the data and interpreting the results. In the long run, a robust photometry methodology with proper nebular subtraction is desirable.

\begin{acknowledgements}
    This work is based on observations obtained with the Samuel Oschin Telescope 48-inch and the 60-inch Telescope at the Palomar Observatory as part of the Zwicky Transient Facility project. ZTF is supported by the National Science Foundation under Grants No. AST-1440341 and AST-2034437 and a collaboration including current partners Caltech, IPAC, the Oskar Klein Center at Stockholm University, the University of Maryland, University of California, Berkeley, the University of Wisconsin at Milwaukee, University of Warwick, Ruhr University Bochum, Cornell University, Northwestern University, and Drexel University. Operations are conducted by COO, IPAC, and UW.

    This work has made use of data from the European Space Agency (ESA) mission \emph{Gaia} (\url{https://www.cosmos.esa.int/gaia}), processed by the \emph{Gaia} Data Processing and Analysis Consortium (DPAC; \url{https://www.cosmos.esa.int/web/gaia/dpac/consortium}). Funding for the DPAC has been provided by national institutions, in particular, the institutions participating in the \emph{Gaia} Multilateral Agreement.

    We are grateful to the staffs of Palomar Observatory and the Hobby-Eberly Telescope for assistance with the observations and data management. The Liverpool Telescope is operated on the island of La Palma by Liverpool John Moores University in the Spanish Observatorio del Roque de los Muchachos of the Instituto de Astrofisica de Canarias with financial support from the UK Science and Technology Facilities Council.

    The Low-Resolution Spectrograph 2 (LRS2) on HET was developed and funded by the University of Texas at Austin McDonald Observatory and Department of Astronomy, and by Pennsylvania State University. We thank the Leibniz-Institut f\"ur Astrophysik Potsdam (AIP) and the Institut f\"ur Astrophysik G\"ottingen (IAG) for their contributions to the construction of the integral field units.
We acknowledge the Texas Advanced Computing Center (TACC) at The University of Texas at Austin for providing high performance computing, visualization, and storage resources that have contributed to the results reported within this paper.

    We thank the anonymous referee for the detailed comments, which improved the clarity of the manuscript significantly. S.B. expresses gratitude to Kishalay De for providing the Gattini-IR and WISE data. S.B. thanks Frank J. Masci and Zachary P. Vanderbosch for useful discussions and suggestions regarding solving the issues with ZTF forced photometry on extended sources. S.B. also thanks Jim Fuller, Charles C. Steidel, Lynne Hillenbrand, and Adolfo Carvalho for useful discussions on methods and science. S.B. also thanks David O. Cook for providing access to his CLU image cutout service to generate the WeSb~1 image. S.B. acknowledges the financial support from the Wallace L. W. Sargent Graduate Fellowship during the first year of his graduate studies at Caltech. N.C. was supported through the Cancer Research UK grant A24042. S.B. thanks Martina Veresvarka for drawing our attention to the TESS light curves of WeSb~1. 

    We have used \texttt{Python} packages Numpy \citep{harris2020array}, SciPy \citep{2020SciPy-NMeth}, Matplotlib \citep{Hunter:2007}, Pandas \citep{reback2020pandas}, Astropy \citep{Astropy13, Astropy18}, and Astroquery \citep{astroquery19} at various stages of this research.
\end{acknowledgements}


\appendix

\section{ZTF Data and the NEV Metric}\label{app:stat}

\begin{figure}[!ht]
\figurenum{A1}
    \centering
    \includegraphics[width=\linewidth]{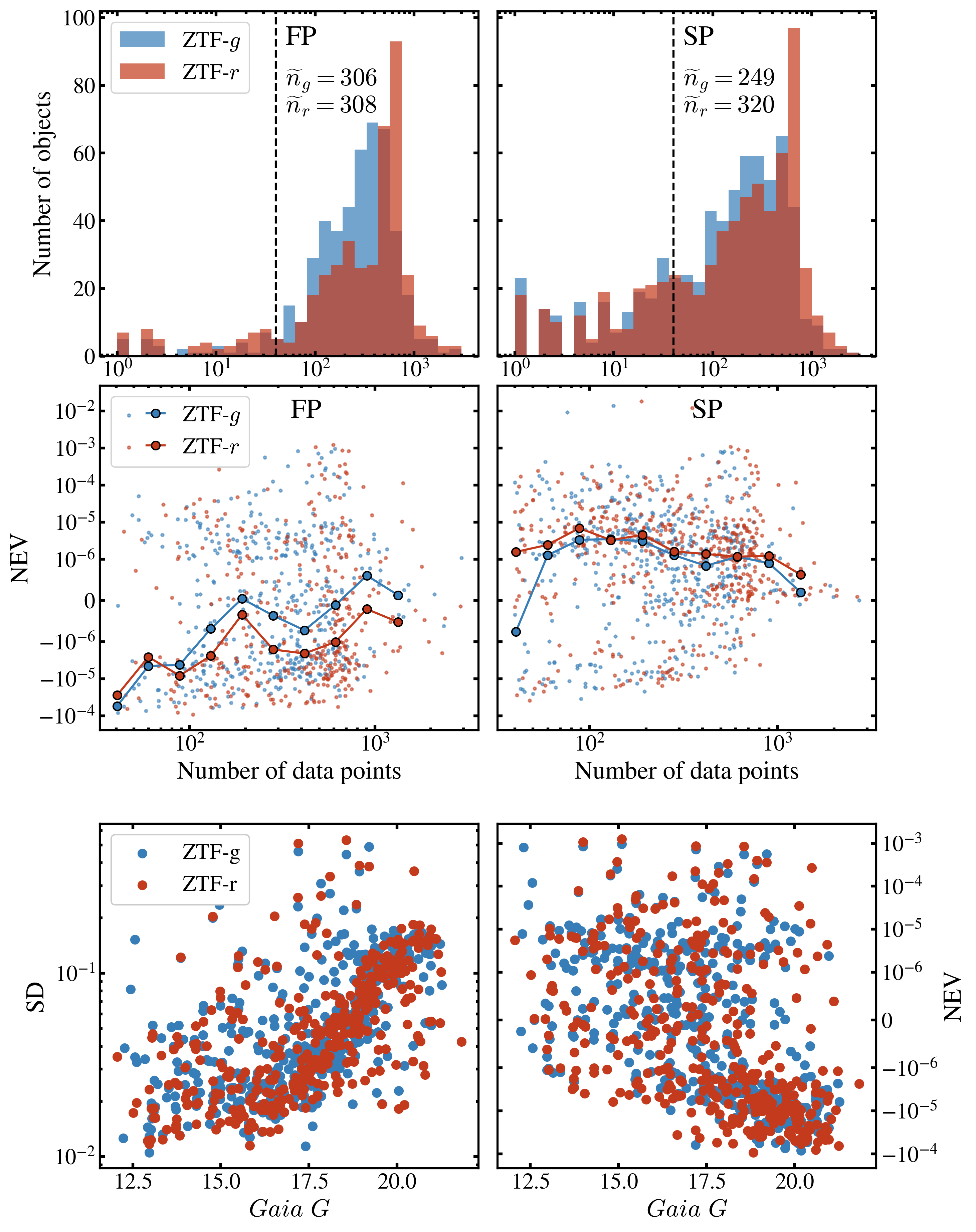}
    \caption{\textbf{Top Panel:} The distribution of the number of good data points for both the ZTF bands and for both forced (FP, left) and standard (SP, right) photometry. The vertical line marks the minimum requirement of 40 good data points. \textbf{Middle Panel:} The NEV metric as a function of the number of data points for all the objects. The solid curve highlights any overall trend in the distribution. \textbf{Bottom Panel:} A comparison of the Standard Deviation (SD, which is a more commonly used metric) with the NEV with regards to their trend with magnitude.}
    \label{fig:number_nev_statv}
\end{figure}

The inclusion of this section is largely motivated from the comments of the anonymous reviewer. Here, we continue our discussion from Section \ref{subsec:metrics_and_period} and further elaborate on the NEV metric and its application to the ZTF data. First, we provide the statistics on the number of data points per object, on which we apply our variability metric. The top panel of Figure \ref{fig:number_nev_statv} shows the distribution of the number of data points for each object which passes all the photometric quality cuts for both the forced photometry (FP) and standard photometry (SP). For both bands, the median number of data points per object is $\approx$$300$, after rejecting all the object with less than $40$ ``good" data points. This shows that, for most objects, we have sufficient number of data points for a reliable evaluation of the variability metric.

The distribution of the number of data points, however, is not narrow. This leads to the question of how much the metric NEV gets artificially affected by the number of data points. In the middle panel of the same figure, we plot the NEV metrics for all the objects as a function of their number of good data points, $N$. The scatter plot shows no correlation. The NEV metric is indeed a normalized metric and does not depend on $N$ (the authors have also verified this through synthetic data). To highlight any trend, we compute the median of the NEVs within logarithmic bins of width $0.16$~dex in $N$. A weak positive correlation is seen with the FP data. However, we reason that this is a ``true" trend and not an artifact. Objects with smaller number of data points are also likely the objects with high uncertainty in photometry (which led to rejection of several of the data points). Thus, they will tend to have lower NEV metric values. With the SP data, however, we do not see the trend. The medians, however, are skewed to higher values. This is due to the underestimation of errors for SP as discussed in Section \ref{subsec:metrics_and_period}.

We note here that, though NEV is not dependent on $N$, it is indeed strongly dependent on the photometric errors. This is one of main reasons why this metric has been chosen, given the often large errors seen in photometry of PNe (see discussion in Section \ref{subsec:photometry}). But this may result in false positives through the uncertainties in the photometric errors themselves. This is prevented through the use of both standard and forced photometry data, which undergo different processing steps and independent error assignments. Thus, an agreement between the two is a strong indication of the true variability of the object.

In the last panel of Figure \ref{fig:number_nev_statv} we justify our choice of NEV over other commonly used metrics like standard deviation (SD). The left (right) panel shows SD (NEV) as a function of the \emph{Gaia}~$G$ magnitude of the objects. We only use FP data for an even comparison. We clearly see that SD suffers from a increasing trend with magnitude, owing to larger photon noise for fainter objects. The trend disappears with NEV. On the contrary, the fainter objects are seen to have lower NEV values. This is also a true trend, as fainter objects are expected to have poorer photometry (and thus larger errors, and lower NEV). This discussion justifies the suitability of NEV as a variability metric for this work.

\section{A unique metric space for unique variables}\label{sec:vonn_skewp}

\begin{figure}[t]
\figurenum{B1}
    \centering
    \includegraphics[width=\linewidth]{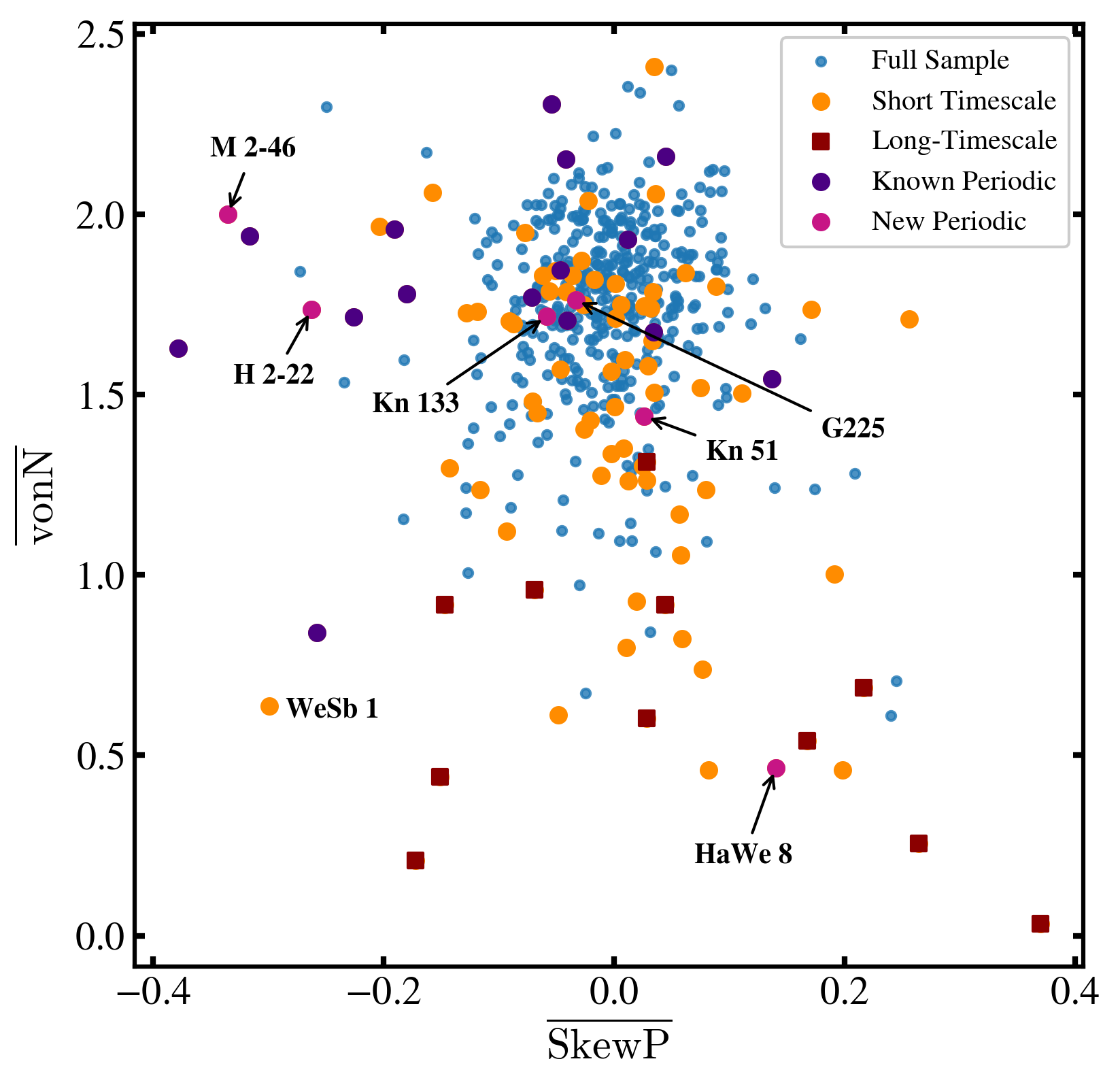}
    \caption{The position of the CSPNe in the 2-dimensional space defined by the variability metrics von Neumann statistics (${\rm vonN}$, Equation \ref{eq:vonn}) and Pearson-Skew (${\rm SkewP}$, Equation \ref{eq:skewp}). The different classes of variables are separately marked (note that `short-timescale' in the legend refers to the aperiodic variables). The long-timescale variables stand out with low ${\rm vonN}$ (further details in Paper II). Several of the short-timescale variables including a few periodic sources and the aperiodic dipper WeSb~1 can be filtered out from the central crowd of non-variable sources. This simple metric space with easy interpretability can be used to identify unique variables of astrophysical interest from a larger pool of objects. See section \ref{sec:vonn_skewp} for more detailed discussion.}
    \label{fig:vonn_skew}
\end{figure}

\begin{figure}[!ht]
\figurenum{C1}
    \centering
    \includegraphics[width=\linewidth]{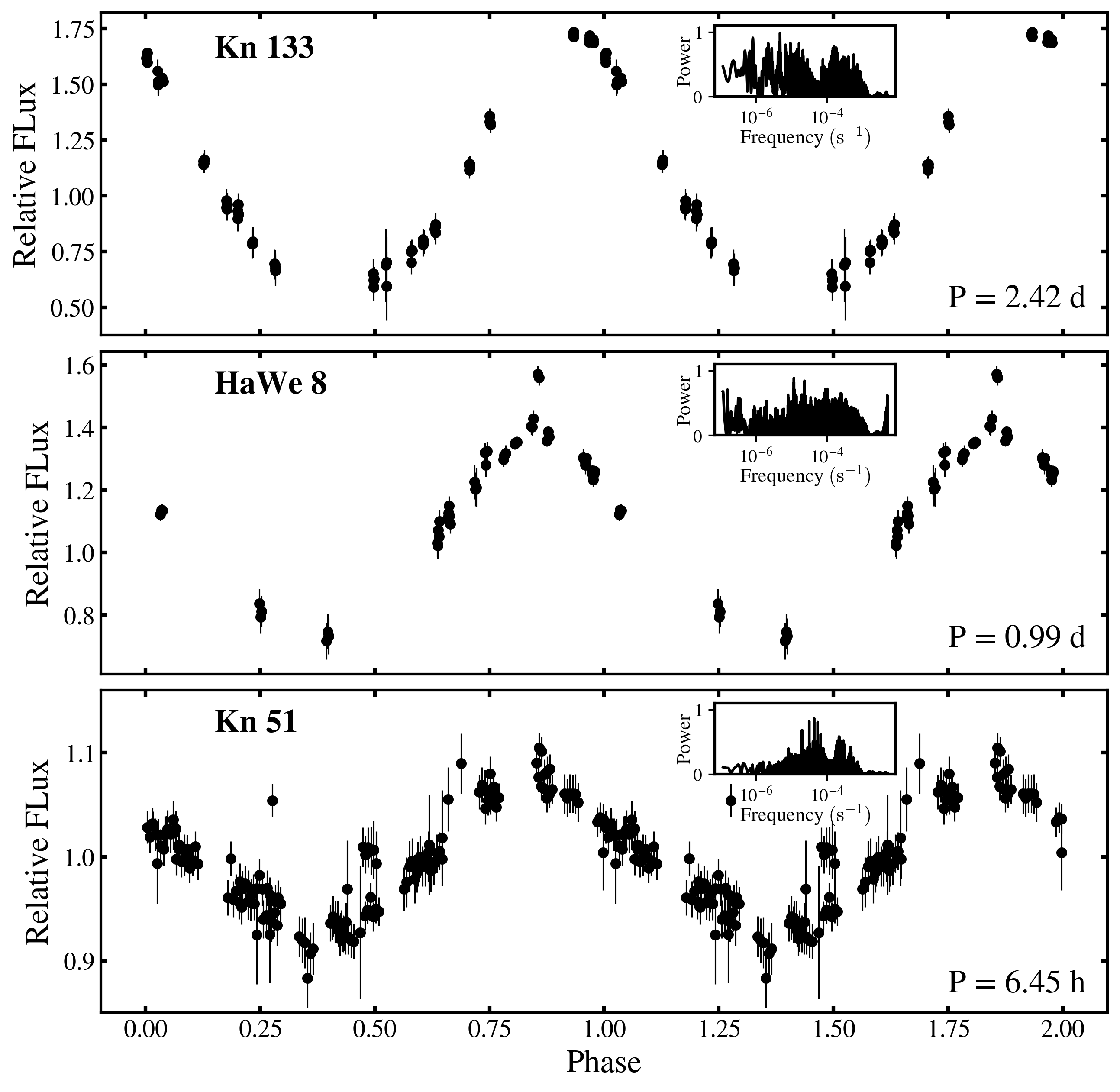}
    \caption{The phase folded follow-up $i$-band light curves for Kn~133, HaWe~8, and Kn~51 with the Liverpool telescope. The inferred periods are well consistent with the ZTF periods. The periodograms are provided as insets. For the phase calculation, the same reference epochs as the ZTF light curves have been used.}
    \label{fig:binary_followups}
\end{figure}

\begin{figure*}[!ht]
\figurenum{D1}
    \centering
    \includegraphics[width=\linewidth]{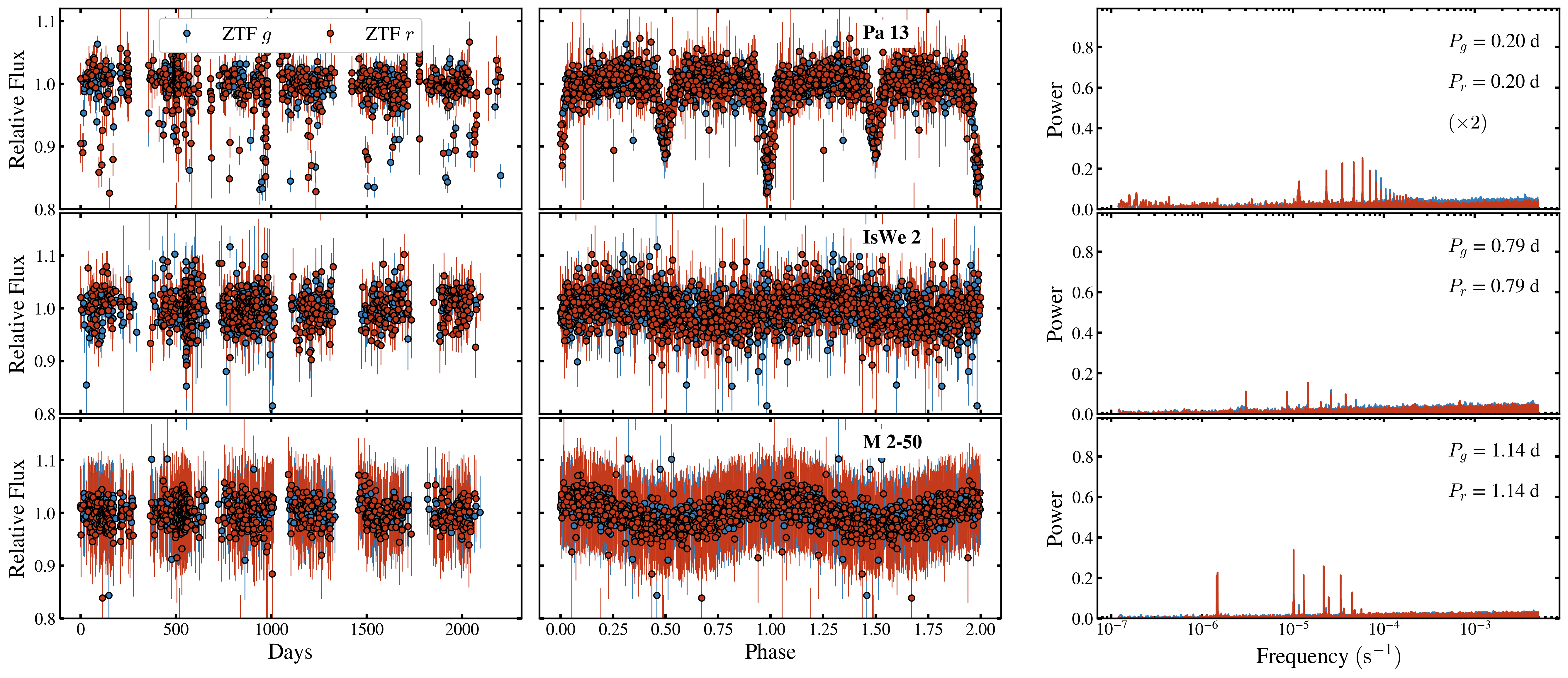}
    \caption{Three new periodic sources among the non-HNEV objects. The figure format is the same as Figure \ref{fig:new_periodics}.}
    \label{fig:new_periodics_non_hnev}
\end{figure*}

In this section, we present the two-dimensional metric space defined by the variability metrics von Neumann (henceforth vonN) and Pearson-Skew (henceforth SkewP) and demonstrate its success in identifying `unique' photometric variables. vonN is defined as \citep{VonNeumann41}:
\begin{equation}\label{eq:vonn}
    \mathrm{vonN} = \left[\frac{\rm RMS(m_{i+1}-m_i)}{\rm SD}\right]^2 = \left[\frac{\mathrm{P2P}}{\rm SD}\right]^2
\end{equation}
where ${\rm SD}$ and $\mathrm{P2P}$ are the standard deviation and point-to-point scatter metrics for the light curves, respectively. The latter is defined as the root-mean-square of the difference of successive data points (in this case, magnitude values) in a time series. SkewP is defined as:
\begin{equation}\label{eq:skewp}
    \mathrm{SkewP} = \frac{\rm median - mean}{\rm SD}.
\end{equation}
vonN quantifies the randomness in the data. It can be shown that the expectation value of vonN for data drawn randomly from a normal distribution is $2$ \citep{VonNeumann41}. Thus, sources with vonN significantly less than $2$ have either long timescale variability well resolved in the given cadence (thus reducing ${\rm P2P}$), or persistent non-random variability. SkewP, on the other hand, quantifies the variability `directionality': ${\rm SkewP}$$>0$ indicates outbursts\slash brightenings, whereas $<$$0$ indicates dips\slash transits. Thus, the selection of objects displaying a certain type of variability can be performed by applying an appropriate cut in this parameter space. This (or similarly defined) metric space has already been used successfully with ZTF data in identifying different populations of sources such as microlensing candidates \citep{Tony21}, young stellar objects \citep{Cody2014,Hillenbrand2022}, and white dwarfs with transiting planetary debris (Bhattacharjee et al. in prep, Vanderbosch et al. in prep). 

We calculate the metric values with only the forced photometry data. Similar to the band-average ${\rm \overline{NEV}}$ metric, we define the mean metric values from the $g$ and $r$ bands as ${\rm \overline{vonN}}$ and ${\rm \overline{SkewP}}$. Figure \ref{fig:vonn_skew} shows the distribution of the objects in this metric space. The median and $3$$\sigma$ trimmed standard deviation values for the ${\rm \overline{vonN}}$ population are $1.8$ and $0.23$ respectively. Note that the median ${\rm \overline{vonN}}$ is broadly consistent with this metric's theoretically expected random value. For ${\rm \overline{SkewP}}$, the median and standard deviation values are $-0.003$ and $0.054$ respectively. These are consistent with the expectation that most objects do not show skewed light curves.

We first note that the long-timescale variables stand out with low vonN metric values, making this metric space particularly effective in identifying such variables. Detailed discussion on these variables will be provided in Paper~II. The positions of the short-timescale variables in this space, however, are less predictable. Jittering sources with temporally unresolved light curves often resemble random data and, thus, result in a high vonN metric. However, even in this regime, sources with high or low SkewP values can be potentially interesting. For example, several of the eclipsing binaries (like M~2-46, candidate source H~2-22, and a few known ones) have large negative SkewP values and are well-separated from the cloud of non-variables. A few sources, with a variability timescale very close to the survey cadence (like HaWe~8, with an inferred period of 0.99 days) can occasionally have low vonN values. Erratically variable sources, like WeSb~1, often tend to have low vonN values too. For this particular object, the partial resolution of the photometric activity (see inset in Figure \ref{fig:wesb1_ztf_gattini_ir}) also contributes to its low Vonn value. The deep dips result in a negative SkewP. These demonstrate the effectiveness of this metric space in identifying `interesting' and `unique' variables based on their low vonN and/or large (positive or negative) SkewP values. Such metric spaces with easy interpretability will be specifically useful in the coming era of large photometric surveys (like LSST).

\section{Follow-up Observations with the Liverpool Telescope}\label{appendix:follow_up_lt}

The Liverpool Telescope observations were obtained  through a Sloan $i$ filter with the IO:O instrument in queue mode under program ID XPL21A08 between October 2021 and February 2022.  The data were reduced by the standard IO:O pipeline and then differential photometry of the central stars was performed against non-variable field stars using standard astropy routines \citep{Astropy18}

We performed an independent Lomb-Scargle period search on the LT data over the same period range and with the same resolution as mentioned in Section \ref{subsec:short_timescale_periodic}. We obtain periods well consistent with those inferred from the ZTF data. This serves as a confirmation for the periods, especially for Kn~51 and HaWe~8, where the ZTF data is noisy and the period is close to 1~day (ZTF cadence), respectively. The phase-folded data are presented in Figure \ref{fig:binary_followups}. For even comparison with the ZTF data, the same epoch has been chosen as the starting phase. The apparent phase shift of the LT data of Kn~51 with respect to the ZTF data, when phase-folded at the LT-inferred period, is an artifact, owing to the small difference ($\sim$2 seconds) in the inferred periods from the two light curves. The difference is unlikely to be real or significant.

\section{Future Prospects: New Periodic Variables among the non-HNEV objects}\label{appendix:future_prospects}

Through the application of strict photometric quality cuts and selection through variability metrics, we have missed several potentially interesting sources appearing in the broader ZTF dataset. To motivate the need for further studies, we present here (Figure \ref{fig:new_periodics_non_hnev}) three new periodic variables which do not appear in the HNEV sample: Pa~13 (PN~G041.4-09.6), IsWe~2 (PN~G107.7+07.8), and M~2-50 (PN~G097.6-02.4). The current HASH catalog lists all three objects as ``True" PNe. To the best of our knowledge, this is the first report of periodicity in these three sources. This readily shows the scope of improvement in the study of CSPNe with ZTF. With regards to periodicity, a more careful study revealing more such sources will be presented in Tam et al. (in preparation).

The light curve of Pa~13 suggests it to be a doubly eclipsing periodic binary system. Based on the visual inspection of the phase-folded light curves, we infer that the Lomb-Scargle period is half the actual period of 9.6 hours. In IsWe~2, we detect very low-amplitude periodic variability with an inferred period of 0.79~days (independently in the $g$ and $r$ bands). Neither of these objects made it to the HNEV cut-off due to the large photometric errors (thus, low NEV metric value). A significant period at 1.14~days is recovered in both ZTF with M~2-50. This source also shows substantial photometric errors (as seen from the Figure). This is also an example of an object that was completely rejected owing to different reference fluxes from the two fields (second issue as discussed in Section \ref{subsec:photometry}). The light curve provided in the figure is from a single field. This also demonstrates that the workaround solutions employed to tackle the photometric issues (see Section \ref{subsec:photometry}) are not optimum and further work is needed on this front.

\section{Spectrum of PN~G030.8+03.4a}\label{appendix:png030spectrum}

\begin{figure*}[!ht]
\figurenum{E1}
    \centering
    \includegraphics[width=\linewidth]{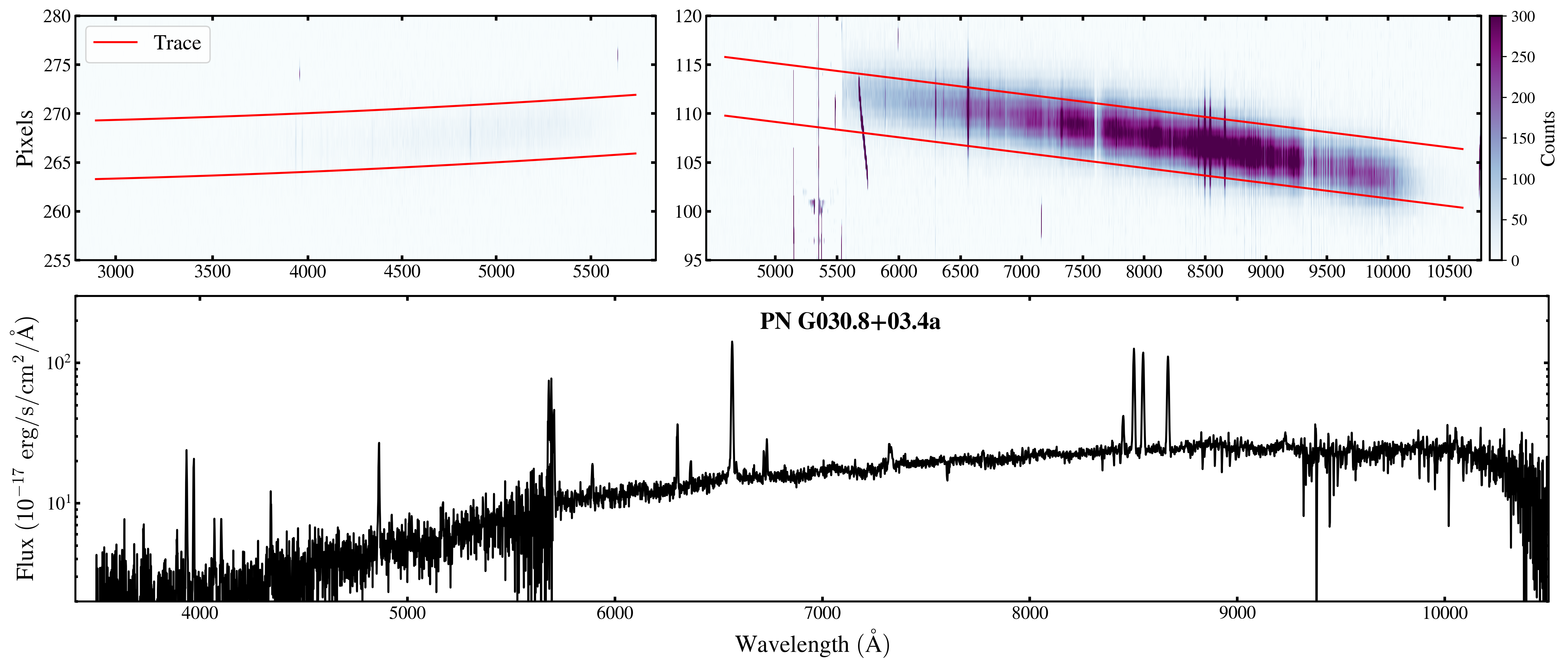}
    \caption{Spectrum of PN~G030.8+03.4a, same format as Figure \ref{fig:WeSb1_trace_spectrum}.}
    \label{fig:G030_trace_spectrum}
\end{figure*}

Optical spectroscopic observation of PN~G030.8+03.4 (one of the objects showing outbursts in the ZTF light curve, see Figure \ref{fig:new_short_vars}) was performed by DBSP on the same date with the same instrument configuration as WeSb~1 (see Section \ref{subsubsec:wesb1_obs}). The same reduction procedure was also followed. Figure \ref{fig:G030_trace_spectrum} shows the spectrum for the object. The unusual feature in the spectra are the strong \ion{Ca}{2}~(H, K, and triplet) emission lines, which are not usually seen in PNe. The spectrum also lacks the [\ion{O}{3}] emission lines. Considering the light curve and lack of nebula, the spectrum is indicative of either a symbiotic system or a CV. We also see prominent [\ion{S}{2}]~6716,~6731~\AA\ emissions in the spectrum. In CVs, such forbidden lines appear when the system has sufficiently evolved since the nova eruption (old novae, see for example the discussion in \citealt{Inight23}). Further observations with better signal-to-noise are needed to confirm the nature of the object, but it is unlikely to be a PN.

\begin{figure*}[!ht]
\figurenum{G1}
    \centering
    \includegraphics[width=\linewidth]{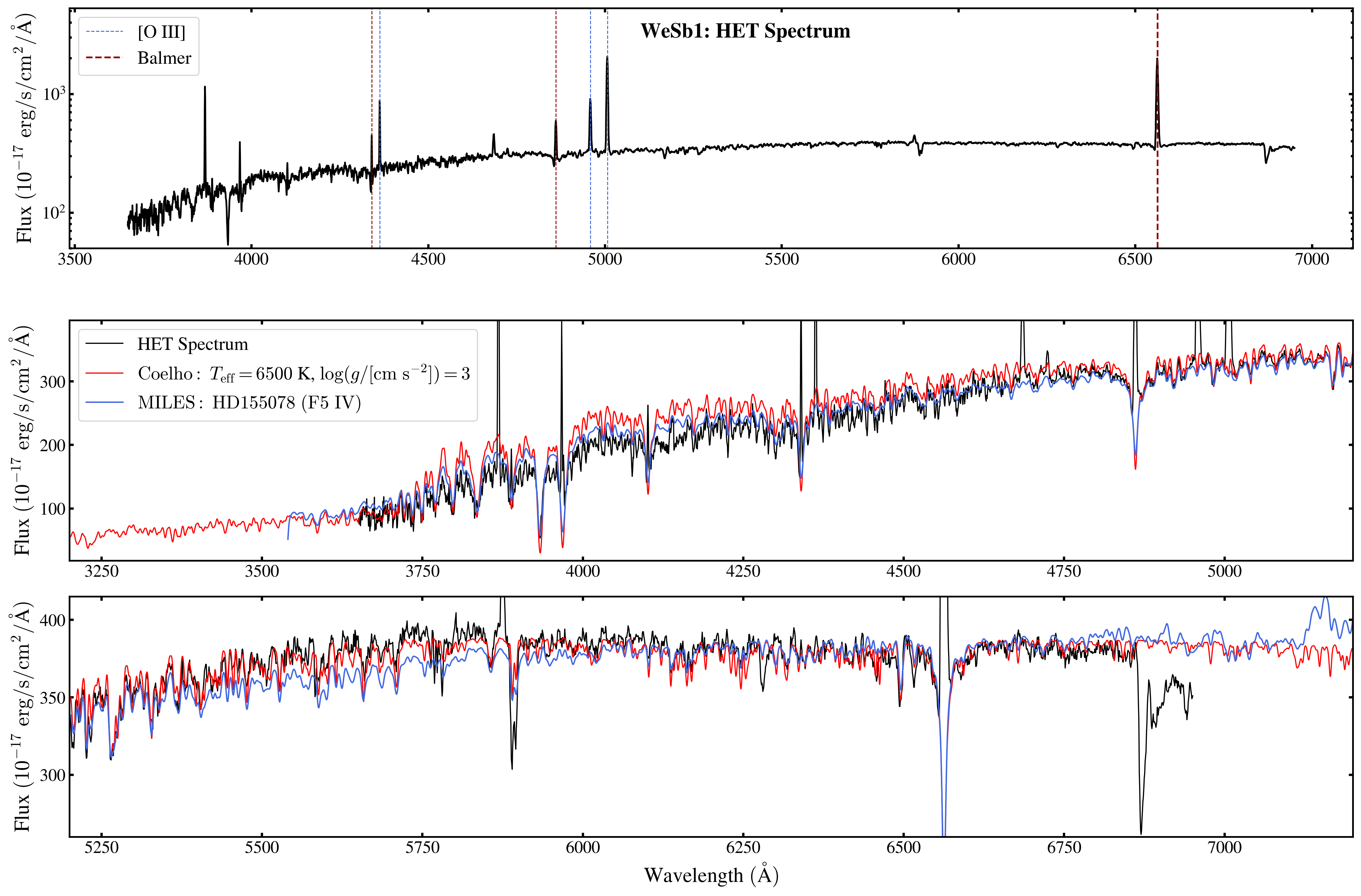}
    \caption{\textbf{Top Panel:} HET spectrum of the central star of WeSb~1. \textbf{Bottom Panel:} Comparison of the HET spectrum with synthetic \texttt{Coelho} spectrum and the spectrum of the F5~IV star HD\,155078, as presented in section \ref{subsubsec:wesb1_spec_analysis}.}
    \label{fig:WeSb1_spec_compare_HET}
\end{figure*}

\section{Misidentified nucleus of W\MakeLowercase{e}S\MakeLowercase{b}~1 or Chance Alignment?}\label{appendix:misidentified}

We now briefly discuss the possibility that this stellar source is not the true nucleus of WeSb~1. With its high-density compact emission and unusual light curve, the object is very unique. The odds of a chance appearance of such a rare system at the center of a 3$'$ nebula is extremely small. Nevertheless, we performed a brief search for other candidate WeSb~1 central stars in the vicinity of the current nucleus. We queried \emph{Gaia} catalog within 45$''$ of the present nucleus (about half the radius of the nebula). The \emph{Gaia} parallax distances to most of these 26 sources are much larger than the approximate nebular distance (say for example as estimated in \citealt{Frew16}), albeit determined with very low parallax-over-errors (POEs). We then manually set the distances of the stars (with low POE) to the nebular distance and compared their position in the \emph{Gaia} CMD. At this distance, the current nucleus is the brightest, though not the bluest. However, none of the other stars showed immediate promise in becoming a candidate for WeSb~1 nucleus. Furthermore, within this radius, only three objects (other than the current nucleus) have a GALEX NUV counterpart. Two objects were rejected for being a nearby star and for geometrical reasons, respectively. The remaining star (closest to the current nucleus in the southeast direction) seemed the most promising alternative. However, the \emph{Gaia} estimated distance is 5125$\pm$1226~pc, which is just marginally consistent with the nebular distance. Manually placing it at the nebular distance, however, makes it intrinsically much fainter and broadly consistent with a main-sequence star in the \emph{Gaia} CMD. Overall, with our preliminary check, we do not find any obvious evidence to discard the current source being the nucleus of WeSb~1, though further confirmatory studies are needed. Even in such a remote possibility of misidentification, the observed system properties are significantly interesting and unique in themselves, even without the nebular association. 

We now consider the possibility where the candidate companion star is a chance alignment with the GALEX source, which is the true CSPN. In this scenario, the age estimation of the PN will be erroneous. We compute the probability of such a chance alignment. The distance between the GALEX and \emph{Gaia} coordinates is 0.54~arcseconds. The G magnitude of the candidate central star is 14.77. We take a radius of 7~arcminutes and find 65 sources as bright as the current star ($G$$<$$15$). Assuming an uniform space density, a chance alignment of such a star with the GALEX coordinate is $\approx$$10^{-4}$, which is quite low. Additionally, the \emph{Gaia} parallax distance, based on the optical source, is compatible to the nebular distance estimated from H$\alpha$ surface brightness and radius relation (see Section \ref{appendix:wesb1}). This strongly suggests a true association between the optical and the UV sources and the nebula.

A second possibility is the chance alignment of the GALEX source with the optical source, the latter being the true CSPN. This is even less likely as the GALEX source density is sparser (25 sources within 7~arcminutes). Also, a hot object (thus a GALEX source) is expected as the star of the candidate companion's property cannot produce high excitation emission lines like [\ion{O}{3}].

\section{HET Spectrum of W\MakeLowercase{e}S\MakeLowercase{b}~1}\label{appendix:wesb1_het}

The one-dimensional HET spectrum of the nucleus of WeSb~1 is shown in the top panel of Figure \ref{fig:WeSb1_spec_compare_HET}. The same features as in DBSP are recovered, in line with expectations. In the bottom two panels of the same figure, we present the comparison of the HET spectrum with the synthetic \texttt{Coelho} and the reference F5~IV star HD\,155078 spectra (same as Figure \ref{fig:WeSb1_spec_compare}, with the comparison spectra appropriately down-resolved to match the resolution of HET observation). As with DBSP, a good match is observed. 


\section{Possible evolutionary stage of the W\MakeLowercase{e}S\MakeLowercase{b}~1 CSPN}\label{appendix:wesb1_cspn_mist_compare}

\begin{figure}[t]
\figurenum{H1}
    \centering
    \includegraphics[width=\linewidth]{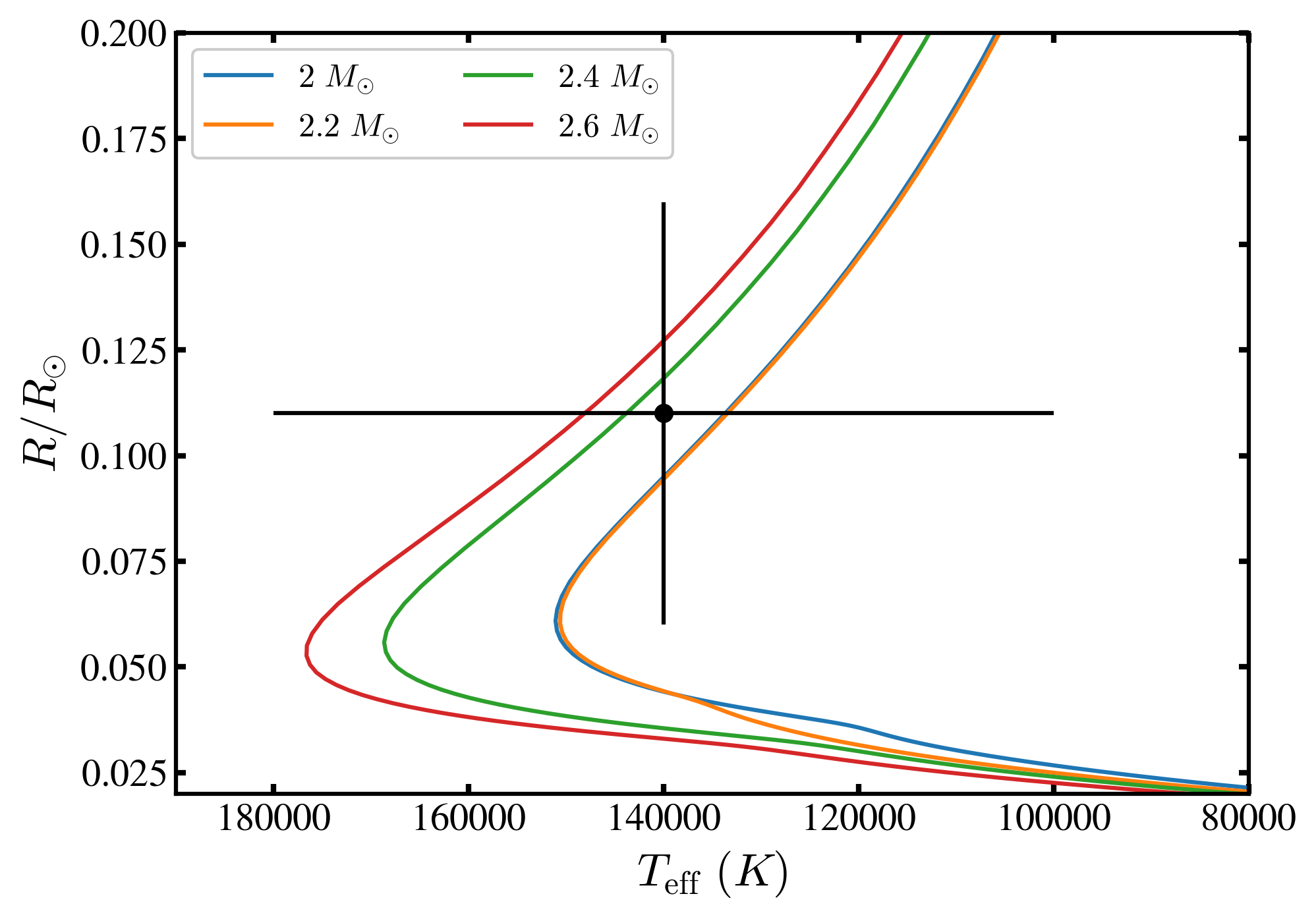}
    \caption{Comparison of the approximate parameters of the CSPN of WeSb~1 (black point) with those predicted by MESA models (colored lines). For the latter, the same four stellar masses as in Figure \ref{fig:wesb1_mist_compare} have been used. The error bars in the CSPN data point are only representative.}
    \label{fig:wesb1_cspn_mist_compare}
\end{figure}

As discussed in section \ref{subsec:wesb1_discussion}, the assumption that the F-type companion is truly in the subgiant phase gives a mass constraint of ${\rm 2.3\pm0.3~M_{\odot}}$. With the PN still visible (thus having evolved off the main sequence just recently), this readily gives a similar mass constraint on the PN progenitor. The current properties of the CSPN could not be well-constrained due to the lack of data points at smaller wavelengths. We take the parameters of the example `fit' in the bottom panel Figure \ref{fig:wesb1_sed} and examine its position in the stellar evolutionary tracks. The comparison with MIST models is presented in Figure \ref{fig:wesb1_cspn_mist_compare} and we indeed see a reasonable agreement. 

We briefly verify that the age of the nebula is consistent with the models. The interval between the epoch of maximum stellar radius (approximately the epoch of the onset of the PN-phase) and the highest temperature of the resultant white dwarf (close to the `inferred' properties of the CSPN of WeSb~1) in the MIST models is ${\rm 2-5\times10^4}$~years. Assuming a radius of ${\rm 1.5}$~pc for the nebula of WeSb~1 (see section \ref{subsec:short_timescale_aperiodic}) and a nominal expansion velocity of ${\rm 50}$~km per second, we get an age estimate of $\sim$$3\times10^4$~years, in very good agreement with the stellar models. Such an age is also reasonable for an evolved PN like WeSb~1.

\section{Known Periodic Variables}

\begin{figure*}[!p]
\figurenum{I1}
    \centering
    \includegraphics[width=\linewidth]{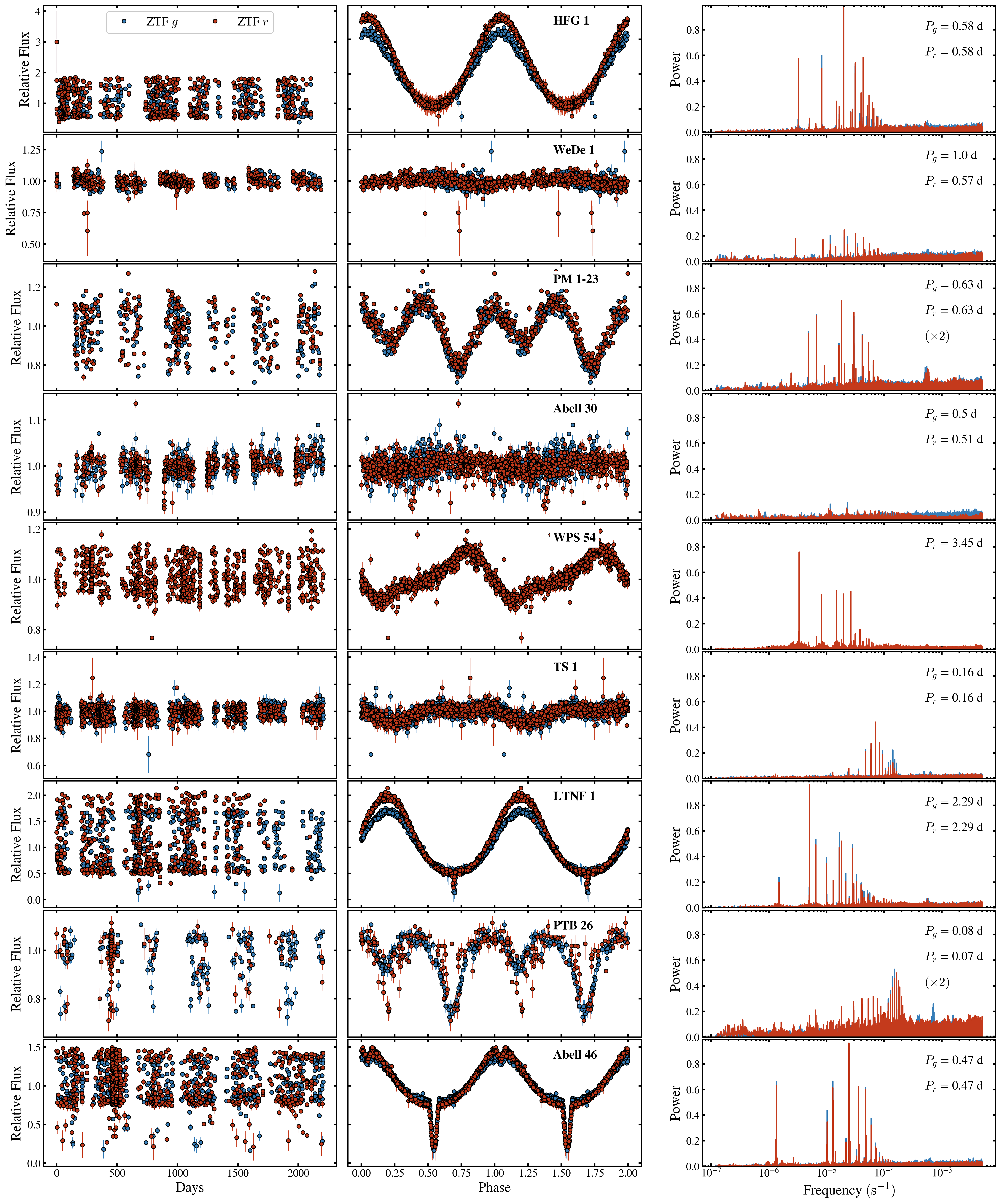}
    \caption{Known periodic CSPNe within the HNEV sample. Same figure format as Figure \ref{fig:new_periodics_non_hnev}}
    \label{fig:known_periodics_1}
\end{figure*}

We present the ZTF light curves (both raw and phase-folded) and the periodograms for the known periodic CSPNe recovered in the HNEV sample in Figure \ref{fig:known_periodics_1}.

\section{List of the short-timescale variables}

We provide the list of the short-timescale HNEV variables in Table \ref{tab:short_timescale_vars}. The non-obvious column names are -- PNStat: the PN status as on the current HASH catalog; PNRad: the PN radius as in the \citetalias{Chornay21Distance} which are derived from those in the HASH catalog; Rel: the reliability score (in the range of 0 -- 1) assigned to the central star identification in \citetalias{Chornay21Distance}.


\begin{figure*}[!p]
\figurenum{I1: continued}
    \centering
    \includegraphics[width=\linewidth]{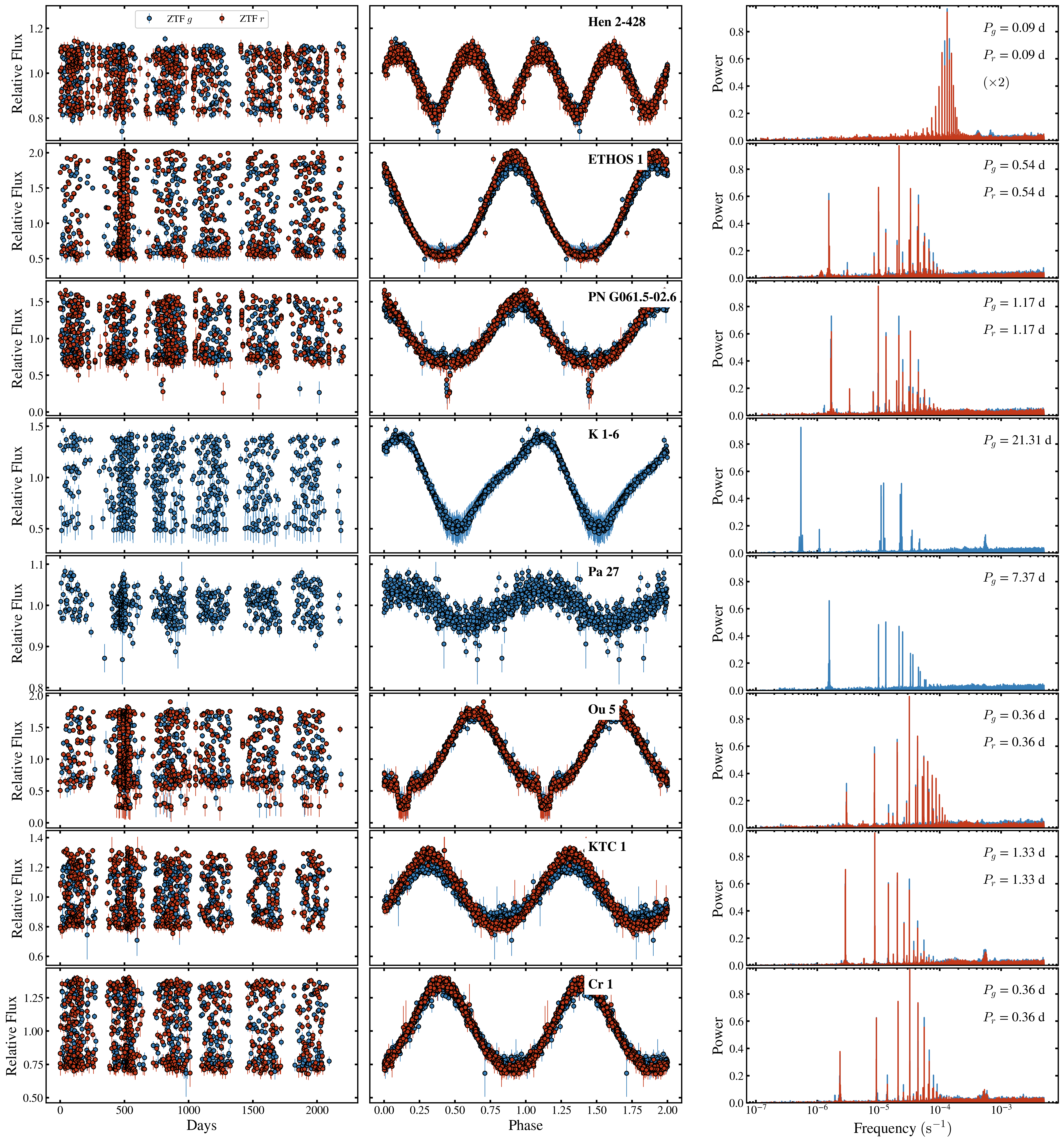}
    \caption{Known periodic CSPNe within the HNEV sample.}
    \label{fig:known_periodics_2}
\end{figure*}

\clearpage

\startlongtable
\centering
\begin{deluxetable*}{ccccccccccc}
\tablenum{J1}
\tablecaption{The short-timescale HNEV variables}
\label{tab:short_timescale_vars}
\tabletypesize{\scriptsize}
\tablewidth{-1pt}
\tablehead{
    \colhead{PNG} &
    \colhead{Name} & 
    \colhead{RA} &  
    \colhead{Dec} & 
    \colhead{GaiaID} & 
    \colhead{PNStat} &
    \colhead{PNRad} &
    \colhead{Rel} &
    \colhead{Known Period} &
    \colhead{ZTF Period} &
    \colhead{Reference}   
}
\startdata
 124.3-07.7  & WeSb 1                    & 15.225  & 55.067  & 423384961080344960  & L      & 92.5   & 1.0  & --          & --        & 1 \\
136.3+05.5  & HFG 1                     & 45.946  & 64.91   & 468033345145186816  & T      & 250.0  & 1.0  & 0.58 d      & 0.58 d    & 10                                      \\
150.9-10.1  & Bode 1                    & 52.8    & 43.904  & 238540495056450048  & P      & 60.0   & 0.99 & --          & --        & --                                      \\
144.1+06.1  & NGC 1501                  & 61.747  & 60.921  & 473712872456844480  & T      & 28.5   & 1.0  & --          & --        & 1,2,8 \\
206.4-40.5  & NGC 1535                  & 63.566  & -12.739 & 3189152962633164800 & T      & 16.65  & 1.0  & --          & --        & --                                      \\
164.8-09.8  & Kn 51                     & 66.362  & 35.102  & 176269718435866112  & T      & 42.0   & 1.0  & --          & 6.45 h        & 1 \\
173.7-09.2  &                           & 73.64   & 28.825  & 155075188702503424  & L      & 15.5   & 1.0  & --          & 2.42 d    & 1 \\
174.5-03.0  & IPHAS J052015.87+314938.8 & 80.066  & 31.827  & 180731845860251264  & P      & 20.0   & 1.0  & --          & --        & 1 \\
197.4-06.4  & WeDe 1                    & 89.854  & 10.695  & 3341996555048041984 & L      & 510.0  & 1.0  & 0.57 d          & 0.57 d ($r$-only)        & 6                                      \\
192.5+07.2  & HaWe 8                    & 100.04  & 21.417  & 3378885131503804416 & T      & 52.5   & 1.0  & --          & 0.99 d    & 1 \\
222.8-04.2  & PM 1-23                   & 103.556 & -10.761 & 3049278041153797632 & T      & 13.5   & 1.0  & 1.26 d      & 0.63 d $(\times 2)$    & 11                                      \\
225.5-02.5  & MPA J0705-1224            & 106.405 & -12.414 & 3045709473089381376 & L      & 10.0   & 1.0  & --          & 2.21 h    & --                                      \\
240.3-07.6  & M 3-2                     & 108.708 & -27.84  & 5609860130542365696 & T      & 6.15   & 0.58 & --          & --        & --                                      \\
232.0+05.7  & SaSt 2-3                  & 117.015 & -14.128 & 3029668728008670208 & T      & 1.25   & 1.0  & --          & --        & --                                      \\
241.0+02.3  & M 3-4                     & 118.797 & -23.637 & 5710725616423348224 & T      & 16.5   & 1.0  & --          & --        & 1 \\
208.5+33.2  & Abell 30                  & 131.723 & 17.88   & 660071056749861888  & T      & 63.5   & 1.0  & 1.06 d          & 1.06 d ($r$-only)        & 7                                      \\
162.1+47.9  & WPS 54                    & 147.858 & 53.159  & 1020641700210987648 & P      & 1800.0 & 1.0  & 3.45 d      & 3.45 d    & 1 \\
135.9+55.9  & TS 1                      & 178.353 & 59.666  & 846615127231002880  & T      & 4.6    & 1.0  & 3.92 h      & 3.92 h    & 11                                      \\
144.8+65.8  & LTNF 1                    & 179.437 & 48.938  & 786919754746647424  & T      & 115.0  & 1.0  & 2.29 d      & 2.29 d    & 12                                      \\
337.9+43.5  & Fr 1-5                    & 218.086 & -12.38  & 6324298665725984768 & P      & 130.0  & 1.0  & --          & --        & 3                                      \\
356.6+07.5  & K 6-20                    & 257.187 & -27.517 & 4108539434533748224 & T      & 6.5    & 0.95 & --          & --        & --                                      \\
358.7+05.2  & M 3-40                    & 260.618 & -27.145 & 4108085748534671872 & T      & 1.1    & 0.96 & --          & --        & --                                      \\
359.8+05.6  & M 2-12                    & 261.006 & -25.99  & 4109870908774509056 & T      & 2.2    & 0.96 & --          & --        & --                                      \\
359.2+04.7  & Th 3-14                   & 261.434 & -26.963 & 4109555589418641408 & T      & 0.85   & 0.65 & --          & --        & --                                      \\
008.3+09.6  & PTB 26                    & 262.305 & -16.795 & 4124562102059801088 & T      & 15.0   & 1.0  & 3.6 h       & 1.8 h $(\times 2)$     & 1 \\
001.3+04.0  & IRAS17301-2538            & 263.305 & -25.673 & 4062070460993015808 & P      & 0.0    & 0.93 & --          & --        & --                                      \\
000.3+02.5  & MPA1736-2715              & 264.124 & -27.256 & 4061311767987964928 & L      & 0.0    & 0.65 & --          & --        & --                                      \\
008.2+06.8  & Hen 2-260                 & 264.739 & -18.293 & 4123264059923436544 & T      & 0.9    & 1.0  & --          & --        & 4                                      \\
008.8+06.2  & CBF 1                     & 265.635 & -18.162 & 4123237018804802048 & T      & 4.05   & 0.92 & --          & --        & --                                      \\
006.3+03.3  & H 2-22                    & 266.891 & -21.79  & 4117062676912301056 & T      & 4.0    & 0.84 & --          & 3.41 h?        & --                                      \\
009.3+04.1  & Th 4-6                    & 267.738 & -18.78  & 4120281909424953344 & T      & 2.0    & 0.93 & --          & --        & --                                      \\
007.4+02.5  & Th 4-8                    & 268.178 & -21.26  & 4070651775691847168 & T      & 4.7    & 0.71 & --          & --        & --                                      \\
011.8+03.7  & PHR J1757-1649            & 269.415 & -16.822 & 4144686669576773632 & P      & 21.0   & 0.99 & --          & --        & 1 \\
015.4+05.6  & Pa 44                     & 269.567 & -12.799 & 4150564551275781120 & L      & 4.5    & 0.98 & --          & --        & --                                      \\
005.1-03.0  & H 1-58                    & 272.308 & -26.041 & 4064833739880077824 & T      & 3.0    & 0.88 & --          & --        & --                                      \\
018.9+03.6  & M 4-8                     & 273.04  & -10.716 & 4157058752294047232 & T      & 2.5    & 0.92 & --          & --        & --                                      \\
012.6-02.7  & M 1-45                    & 275.783 & -19.285 & 4096032936429271040 & T      & 4.5    & 0.93 & --          & --        & --                                      \\
013.8-02.8  & SaWe 3                    & 276.515 & -18.21  & 4096221021593115136 & T      & 55.0   & 0.99 & --          & --        & --                                      \\
016.4-01.9  & M 1-46                    & 276.985 & -15.548 & 4103910524954236928 & T      & 6.05   & 0.96 & --          & --        & --                                      \\
013.4-04.2  & V-V 3-4                   & 277.642 & -19.246 & 4092899019087172608 & T      & 0.0    & 0.9  & --          & --        & --                                      \\
055.4+16.0  & Abell 46                  & 277.827 & 26.937  & 4585381817643702272 & T      & 48.5   & 0.99 & 11.32 h     & 11.32 h   & 13                                      \\
009.4-06.6  & PPA J1831-2356            & 277.985 & -23.935 & 4077538635493652992 & T      & 2.75   & 1.0  & --          & --        & --                                      \\
030.8+03.4a & IPHAS J183531.31-001550.0 & 278.88  & -0.264  & 4272532521447280640 & P      & 0.0    & 0.98 & --          & --        & --                                      \\
011.7-06.6  & M 1-55                    & 279.141 & -21.817 & 4079529404360161792 & T      & 3.0    & 0.98 & --          & --        & --                                      \\
011.7-07.0  & Kn 77                     & 279.507 & -22.047 & 4079460719236665344 & N      & 0.0    & 0.87 & --          & --        & --                                      \\
014.4-06.1  & SB 19                     & 279.917 & -19.237 & 4092512742510443008 & T      & 5.35   & 1.0  & --          & --        & --                                      \\
019.2-04.4  & PM 1-251                  & 280.603 & -14.253 & 4103595858442902016 & T      & 9.25   & 0.65 & --          & --        & --                                      \\
014.2-07.3  & M 3-31                    & 281.007 & -19.915 & 4080221267663409152 & T      & 3.5    & 1.0  & --          & --        & --                                      \\
024.8-02.7  & M 2-46                    & 281.644 & -8.467  & 4251616408769092608 & T      & 2.2    & 0.93 & --          & 7.66 h    & --                                      \\
011.3-09.4  & My 121                    & 281.646 & -23.447 & 4078224382749920768 & T      & 1.5    & 0.99 & --          & --        & --                                      \\
027.9-01.4  & PHR J1847-0507            & 281.869 & -5.123  & 4255077121584499200 & L      & 51.5   & 0.94 & --          & --        & --                                      \\
042.0+05.4  & K 3-14                    & 282.137 & 10.597  & 4503757445981896704 & T      & 0.0    & 1.0  & --          & --        & --                                      \\
020.7-05.9  & Sa 1-8                    & 282.685 & -13.517 & 4102207686291028992 & T      & 4.0    & 0.99 & --          & --        & --                                      \\
025.6-03.6  & PM 1-259                  & 282.766 & -8.122  & 4251974334170635776 & T      & 0.0    & 0.79 & --          & --        & --                                      \\
020.3-06.9  & Pa 120                    & 283.431 & -14.333 & 4101903087213994496 & P      & 11.5   & 1.0  & --          & --        & 1 \\
025.1-04.6  & Pa 121                    & 283.519 & -9.03   & 4203637225993816576 & P      & 0.0    & 1.0  & --          & --        & --                                      \\
043.1+03.8  & M 1-65                    & 284.14  & 10.869  & 4311868924367555072 & T      & 2.1    & 0.89 & --          & --        & --                                      \\
049.4+02.4  & Hen 2-428                 & 288.273 & 15.778  & 4513170261953875968 & T      & 20.0   & 0.76 & 4.22 h      & 2.11 h $(\times 2)$    & 14                                      \\
038.4-03.3  & K 4-19                    & 288.344 & 3.417   & 4268281088010980864 & T      & 0.0    & 0.76 & --          & --        & --                                      \\
068.1+11.0  & ETHOS 1                   & 289.131 & 36.163  & 2050526964622031872 & T      & 9.75   & 1.0  & 0.54 d      & 0.54 d    & 15                                      \\
050.0+01.7  & IPHAS J191701.33+155947.8 & 289.256 & 15.997  & 4321039328043783168 & P      & 0.0    & 0.94 & --          & --        & --                                      \\
039.0-04.0  & IPHASX J191716.4+033447   & 289.318 & 3.58    & 4292267621344389120 & L      & 16.0   & 0.84 & --          & --        & 1 \\
043.2-02.0  & PM 2-40                   & 289.461 & 8.252   & 4308012932750879232 & L      & 0.0    & 0.99 & --          & --        & --                                      \\
046.6-01.4  & Pa 131                    & 290.525 & 11.545  & 4315858566694277632 & P      & 5.0    & 0.82 & --          & --        & --                                      \\
055.6+02.1  & Hen 1-2                   & 291.657 & 21.157  & 2018024881202947072 & T      & 2.5    & 0.95 & --          & --        & --                                      \\
045.7-04.5  & NGC 6804                  & 292.896 & 9.225   & 4296362443149857792 & T      & 29.15  & 1.0  & --          & --        & 1 \\
059.9+02.0  & K 3-39                    & 293.977 & 24.914  & 2021426048629219072 & T      & 3.5    & 0.96 & --          & --        & --                                      \\
034.5-11.7  & PM 1-308                  & 294.073 & -3.89   & 4210278482327706624 & T      & 0.9    & 1.0  & --          & --        & 5                                      \\
064.5+03.4  & Kn 15                     & 295.168 & 29.503  & 2031657180462828288 & T      & 13.15  & 1.0  & --          & --        & 1 \\
054.4-02.5  & M 1-72                    & 295.392 & 17.755  & 1824067323576290304 & T      & 2.0    & 1.0  & --          & --        & --                                      \\
026.1-17.6  & Pa 161                    & 295.869 & -13.75  & 4183333506776770560 & T      & 262.5  & 1.0  & --          & --        & 1 \\
066.8+02.9  & IPHASX J194751.9+311818   & 296.966 & 31.305  & 2033745393571783936 & T      & 6.45   & 0.99 & --          & --        & --                                      \\
079.8+08.6  & Kn 27                     & 299.004 & 45.388  & 2079241535416132096 & P      & 0.0    & 0.87 & --          & --        & --                                      \\
061.5-02.6  &                           & 299.347 & 23.88   & 1834171384397003520 & L      & 52.5   & 0.7  & 1.17 d      & 1.17 d    & 1 \\
070.0+01.8a & IRAS 19581+3320           & 300.029 & 33.484  & 2034409361147890432 & P      & 0.0    & 0.92 & --          & --        & --                                      \\
070.9+02.2  & PM 1-316                  & 300.22  & 34.473  & 2058539140238094336 & P      & 0.0    & 0.98 & --          & --        & --                                      \\
107.0+21.3  & K 1-6                     & 301.059 & 74.427  & 2288467186442571008 & T      & 99.0   & 1.0  & 21.31 d     & 21.31 d   & 16                                      \\
075.0-07.2  & Pa 27                     & 312.243 & 32.304  & 1859955657931121664 & T      & 36.0   & 1.0  & --          & 7.37 d    & 1,9 \\
086.9-03.4  & Ou 5                      & 318.583 & 43.693  & 1970016153397634048 & T      & 16.5   & 1.0  & 8.74 h      & 8.74 h    & 13                                      \\
099.1+05.7  & KTC 1                     & 322.046 & 58.876  & 2179544655458448384 & T      & 11.0   & 0.91 & 1.33 d      & 1.33 d    & 1 \\
093.9-00.1  & IRAS 21282+5050           & 322.494 & 51.067  & 2171652769005709568 & T      & 3.0    & 0.7  & --          & --        & --                                      \\
100.3+02.8  & Cr 1                      & 327.299 & 57.455  & 2202260634408051968 & P      & 60.0   & 1.0  & 8.57 h      & 8.57 h    & 1 \\
111.2+07.0  & KjPn 6                    & 342.259 & 67.027  & 2212671428967795968 & T      & 3.0    & 0.96 & --          & --        & --                                                           
\enddata
\tablerefs{1) \cite{Chornay21Binary}, 2) \cite{Bond96}, 3) \cite{Nagel04}, 4) \cite{Hajduk14}, 5) \cite{Arkhipova12}, 6) \cite{Reindl21}, 7) \cite{Jacoby20Abell30}, 8) \cite{Aller24}, 9) \cite{Bond24}, 10) \cite{Exter05}, 10) \cite{Hajduk10}, 11) \cite{Tovmassian04}, 12) \cite{Shimanski08}, 13) \cite{Corradi15}, 14) \cite{Santander15}, 15) \cite{Munday20}, 16) \cite{Frew11}}

\end{deluxetable*}


\bibliography{sample631}{}
\bibliographystyle{aasjournal}



\end{document}